\documentclass[11pt,x11names,a4paper]{article}
\usepackage[UKenglish]{babel}
\usepackage{setspace}
\usepackage{titlesec} 	% Required for manipulating the titles
\usepackage{titletoc} 	% Required for manipulating the table of contents
\usepackage[margin=1.5em,labelfont=bf]{caption}
\usepackage{enumitem}
\usepackage{multirow}
\usepackage{float}
\usepackage[section]{placeins}	% Keeps figures in their section. Puts \FloatBarrier at the end of each section
\usepackage{booktabs}
\usepackage{cite}
\usepackage{hyperref}	% hyperref settings
\hypersetup{
	colorlinks=true,
	linkcolor=Blue4,
	citecolor=Red4,
	urlcolor=Green4,
	linktoc=all
}

\usepackage[utf8]{inputenc}
\usepackage{lmodern}
\usepackage{microtype} 		% Slightly tweak font spacing for aesthetics
\usepackage[T1]{fontenc} 	% Use 8-bit encoding that has 256 glyphs
\usepackage{mathrsfs}		% Provides the \mathscr font
\usepackage{amsmath,amscd,amsthm,amssymb} % Do not add amssymb.together with mathdesign
\usepackage{mathtools}
\usepackage{esint,slashed}
\usepackage{graphicx,xcolor,xparse}
\usepackage{tikz} 
\usepackage{tikz-cd} 
\usetikzlibrary{matrix,calc} 
\usetikzlibrary{decorations.markings,shapes.geometric,shapes.misc} 
\usetikzlibrary{decorations.pathreplacing,positioning} 
\usetikzlibrary{arrows}

\usepackage[a4paper, left=25mm, right=25mm, top=30mm, bottom=25mm]{geometry} % Document margins
\numberwithin{equation}{section} 

\titleformat{\section}[block]{\Large\bfseries\centering}{\thesection}{1em}{} 
\titleformat{\subsection}[block]{\bfseries}{\thesubsection}{1em}{} 

\titlespacing*{\section}{0pt}{1em}{1em}
\titlespacing*{\subsection}{0pt}{0.75em}{0.75em}

\addto\captionsUKenglish{
	\renewcommand{\contentsname}%
	{Table of Contents}%
}

\usepackage{abstract}

\def\mop#1{\mathop{\rm #1}\nolimits}

\def\coth{\mop{coth}}

\def\diag{\mop{diag}}
\def\ii{{\rm i}}

\def\AdS{\mop{AdS}}
\def\dS{\mop{dS}}
\newcommand{\ds}{{\rm d}s}
\newcommand{\dd}{{\rm d}}

\newcommand{\e}{\mathrm{e}}

\newcommand\Tr{\mathrm{Tr}\,}

\newcommand{\dvol}{{\rm vol}}
\newcommand{\vol}{\mathbf{V}}
\newcommand{\sgn}{\mathrm{sgn}}

\newcommand{\f}[2]{\frac{#1}{#2}}
\newcommand{\comm}[2]{\left[{#1},{#2}\right]}

\newcommand{\vev}[1]{\left\langle{#1}\right\rangle}

\newcommand{\wti}[1]{\widetilde{#1}}

\newcommand{\disc}{\mathbb{D}}

\newcommand{\be}{\begin{equation}}
	\newcommand{\ee}{\end{equation}}
\newcommand {\bes} {\begin {equation*}}
\newcommand {\ees} {\end {equation*}}
\newcommand{\bea}{\begin{eqnarray*}}
	\newcommand{\eea}{\end{eqnarray*}}

\newcommand{\SU}{\mop{SU}}
\newcommand{\SO}{\mop{SO}}
\newcommand{\SL}{\mop{SL}}

\renewcommand\sl{\mathfrak{sl}}

\newcommand\so{\mathfrak{so}}
\newcommand\su{\mathfrak{su}}

\newcommand\bC{\mathbf{C}}

\newcommand\bU{\mathbf{U}}
\newcommand\cA{\mathcal{A}}

\newcommand\cD{\mathcal{D}}

\newcommand\cH{\mathcal{H}}
\newcommand\cI{\mathcal{I}}

\newcommand\cK{\mathcal{K}}
\newcommand\cL{\mathcal{L}}
\newcommand\cM{\mathcal{M}}

\newcommand\cO{\mathcal{O}}

\newcommand\cR{\mathcal{R}}
\newcommand\cS{\mathcal{S}}

\newcommand\cV{\mathcal{V}}
\newcommand\cW{\mathcal{W}}

\graphicspath{{./figures/}}

\title{\fontsize{20pt}{24pt}\selectfont\textbf{Spherical Branes and the BMN Matrix Quantum Mechanics}\vspace{5mm}}

\author{
	\large{
		\href{mailto:nikolay.bobev@kuleuven.be}{Nikolay Bobev}$^{1,2}$, \href{mailto:pieter.bomans@maths.ox.ac.uk}{Pieter Bomans}$^{3}$,
		\href{mailto:ffg1m24@soton.ac.uk}{Fri{\dh}rik Freyr Gautason}$^{4,5}$, }
	\\[5mm]
	\normalsize $^1$ Instituut voor Theoretische Fysica \& Leuven Gravity Institute, KU Leuven\\
	\normalsize Celestijnenlaan 200D, 3000 Leuven, Belgium\\[5mm]
	\normalsize $^2$ School of Natural Sciences, Institute for Advanced Study\\
	\normalsize 1 Einstein Drive, Princeton, NJ 08540, USA\\[5mm]
	\normalsize $^3\,$Mathematical Institute, University of Oxford\\
	\normalsize Andrew Wiles Building, Radcliffe Observatory Quarter,\\
	\normalsize Woodstock Road, Oxford, OX2 6GG, U.K.\\[5mm]
	\normalsize $^4\,$STAG Research centre \& Mathematical Sciences, University of Southampton\\
	\normalsize Highfield, Southampton SO17 1BJ, U.K.\\[5mm]
	\normalsize $^5$ Science Institute, University of Iceland\\
	\normalsize Dunhaga 3, 107 Reykjav{\'i}k, Iceland
}

\date{}

\begin{document}  
	
{
	\hypersetup{urlcolor=black}
	\maketitle
}

\thispagestyle{empty}	
	
\vspace{\stretch{1}}
		
\begin{abstract}
	\noindent We study the maximally supersymmetric Yang-Mills theory on $S^d$ using supersymmetric localisation and holography. We argue that the analytic continuation in dimension to $d=1$ yields a Euclidean version of the BMN matrix quantum mechanics. This system can be analysed at large $N$ using supersymmetric localisation and leads to explicit results for the free energy on $S^d$ and the expectation value of supersymmetric Wilson loops. We show how these results can be reproduced at strong gauge coupling using holography by employing spherical D-brane solutions. We construct these solutions for any value of $d$ using an effective supergravity description and pay particular attention to the subtleties arising in the $d\to1$ limit. Our results have implications for the BMN matrix quantum mechanics and the physics of circular D0-branes.

\end{abstract}
	
\vspace{\stretch{3}}
	
\newpage
\thispagestyle{empty}

{
	\singlespacing
	\hypersetup{linkcolor=black}
	\setcounter{tocdepth}{2}
	\tableofcontents
}
	
%\newpage
\setcounter{page}{1}

\bigskip
\bigskip
\bigskip	
	
%%%%%%%%%%%%%%%%%%%%%%%%%%%%%%%%%
\section{Introduction}          
\label{sec:introduction}	    
%%%%%%%%%%%%%%%%%%%%%%%%%%%%%%%%%

More than 25 years after the first concrete manifestation of the gauge/gravity duality in string theory, the most understood and widely explored examples are still limited to QFTs with conformal symmetry dual to AdS backgrounds of string and M-theory. The strongly interacting gauge theories arising on the worldvolume of D$p$-branes are in general non-conformal and, as suggested long ago in \cite{Itzhaki:1998dd}, should belong to the broader set of gauge/gravity dualities in string theory. Unfortunately, even in the presence of maximal supersymmetry, there are few quantitative tools available to study holography in these ``non-conformal'' brane setups both in QFT and in string theory and supergravity. This work is a modest attempt to improve the status quo and build on \cite{Bobev:2018ugk,Bobev:2019bvq} to understand better holography for spherical D$p$-branes with a particular focus on the case of D0-branes.

The premise of \cite{Bobev:2018ugk,Bobev:2019bvq} is to employ the fact that maximally supersymmetric Yang-Mills theory (MSYM) can be defined on $S^d$ for $d\leq 7$ while preserving the maximal number of 16 real supercharges \cite{Blau:2000xg}. This in turn suggests that there should be supersymmetric Euclidean D$p$-brane configurations, $d=p+1$, which realise this MSYM theory on their $S^d$ worldvolume. While, it is not immediately clear how to describe these spherical D-branes in open string theory, it was shown in \cite{Bobev:2018ugk,Bobev:2019bvq} how to construct supersymmetric type II supergravity solutions that describe the backreaction of a large number $N$ of such coincident spherical branes. These supergravity solutions are asymptotic to the backgrounds in \cite{Itzhaki:1998dd} describing the near horizon limit of coincident D-branes with flat worldvolumes and can be studied holographically employing the detailed holographic renormalisation procedure developed in \cite{Kanitscheider:2008kd}. On the field theory side of the duality, placing the MSYM theory on $S^d$ is advantageous since one can use the supersymmetric localisation results of  \cite{Minahan:2015jta,Minahan:2015any} to explicitly calculate some observables in the theory in the strong coupling regime relevant to supergravity. As shown in \cite{Bobev:2019bvq}, this program can be brought to bear, allowing one to calculate the $S^d$ free energy and the vacuum expectation value (vev) of the $\frac{1}{2}$-BPS Wilson loop wrapping the equator of $S^d$ at large $N$ and large 't Hooft coupling $\lambda$ using the supersymmetric localisation matrix model of \cite{Minahan:2015jta,Minahan:2015any}. For integer values of $d$ in the range $2 \leq d \leq 7$ these QFT results agree exactly with the holographic analysis performed using the spherical brane solutions in \cite{Bobev:2018ugk}. As emphasised recently in \cite{Biggs:2023sqw}, see also \cite{Boonstra:1998mp,Smilga:2008bt,Wiseman:2013cda} for previous work, the non-conformal MSYM theories on the worldvolumes of D-branes enjoy a scaling similarity in the large $N$ and strong coupling regime which is dictated by the properties of type II supergravity. Indeed, the supersymmetric localisation and holography results of \cite{Bobev:2018ugk,Bobev:2019bvq} explicitly exhibit this scaling similarity in the context of spherical branes. 

Given these developments, it is natural to wonder whether similar calculations can be performed for the two types of coincident D-branes not analysed in \cite{Bobev:2018ugk,Bobev:2019bvq}, namely D0- and D$(-1)$-branes. A primary goal of this work is to shed light on this question for circular D0-branes.\footnote{Understanding D$(-1)$-branes with our methods appear to be more subtle and will only be discussed briefly.} From the perspective of MSYM theory on $S^d$ D0-branes lead to several subtleties. The construction of the MSYM Lagrangian on $S^d$ of radius $\mathcal{R}$ necessitates the addition of new interaction terms proportional to $1/\mathcal{R}$ that are induced by the curvature on the sphere \cite{Blau:2000xg}. These interactions break the $\so(1,9-d)$ R-symmetry of the theory on flat $\mathbf{R}^d$ to its $\su(1,1)\times \so(7-d)$ subalgebra, while preserving 16 supercharges.  On the other hand, if one is interested in the $d=1$ MSYM theory one may just proceed by reducing the 10d $\mathcal{N}=1$ SYM theory on $T^9$ and then find the BFSS matrix quantum mechanics Lagrangian with $\so(9)$ R-symmetry  \cite{Banks:1996vh}. Indeed, as we discuss in detail in Section~\ref{sec:BMN}, there appear to be several inequivalent ways to obtain a Euclidean 1d MSYM theory. One is the torus dimensional reduction that leads to the Euclidean BFSS theory with $\so(1,8)$ R-symmetry. Another one is via analytic continuation of the Euclidean Lagrangian of the MSYM theory on $\mathbf{H}_d$ to $d=1$ which leads to an action with $\su(2)\times \so(1,5)$ R-symmetry as described by Blau in \cite{Blau:2000xg}. Yet another approach, which we will focus on in this work, is  to analytically continue the MSYM Lagrangian on $S^d$ to $d=1$ and obtain a Lagrangian with $\su(1,1)\times \so(6)$ R-symmetry. These last two procedures appear to be, perhaps not surprisingly, closely related to the BMN deformation of the BFSS matrix quantum mechanics which preserves the maximal number of 16 supercharges and has $\su(2) \times \so(6)$ R-symmetry \cite{Berenstein:2002jq}.

The Euclidean BMN matrix quantum mechanics with $\su(1,1)\times \so(6)$ R-symmetry obtained by analytic continuation of the MSYM theory on $S^d$ to $d=1$ appears to have different physics from its well-known Lorentzian avatar. For instance, we find that this maximally supersymmetric model has no supersymmetric vacua with condensing scalars. This is very different from the rich collection of supersymmetric vacua of the Lorentzian BMN model \cite{Berenstein:2002jq} which can be interpreted in terms of polarised D-branes \cite{Myers:1999ps,Maldacena:2002rb} and lead to an equally rich set of supergravity solutions as explained in  \cite{Lin:2005nh}. The Euclidean model on $S^1$ however has the benefit that it can be studied using supersymmetric localisation by analytically continuing the $S^d$ results of \cite{Minahan:2015jta,Minahan:2015any,Bobev:2019bvq} to $d=1$. As shown in \cite{Bobev:2019bvq} the supersymmetric localisation results simplify in the large $N$ limit and when the dimensionless 't~Hooft coupling $\lambda = g_\text{YM}^2 N \cR^{4-d}$ is large one can find explicit expressions for the $S^d$ free energy and the vev of the $\frac{1}{2}$-BPS circular Wilson loop. Using these explicit results and taking $d=1+\epsilon$ leads to the following singular expressions in the $\epsilon \to 0$ limit
\begin{equation}
	\label{eq:FlogWintro}
		F_{1+\epsilon} =\, -\f{5\pi N^2}{7}\bigg(\f{200\epsilon^3}{3\lambda^3}\bigg)^{1/5}\,,\qquad	\log \vev{W_{1+\epsilon}} =\, \pi \bigg( \f{720\lambda}{\epsilon} \bigg)^{1/5}\,.	
\end{equation}
While these expressions have the correct scaling with $N$ and the expected scaling similar behaviour in $\lambda$, see \cite{Biggs:2023sqw}, they clearly need to be regularised. This can be done in an ad hoc manner by rescaling the 't Hooft coupling, or by considering combinations of the two observables that are finite. For instance, the following combination
\begin{equation}
	\label{eq:FW3-prediction}
	F_{1+\epsilon} (\log \vev{W_{1+\epsilon}})^3 = -\f{600\pi^{4}N^2}{7}\,,
\end{equation}
is finite in the $\epsilon \to 0$ limit and provides a convenient target for a holographic analysis in the supergravity limit.

Indeed, turning to supergravity, we find that the singular behaviour of the QFT observables~\eqref{eq:FlogWintro} is mirrored by subtle singularities in the supergravity description of  circular D0-branes. To exhibit this we adopt the approach followed in \cite{Bobev:2018ugk} and study a consistent truncation of 2d maximal gauged supergravity with $\su(1,1)\times \so(6)$ symmetry, which in turn is a consistent truncation of type IIA supergravity, that consists of the metric and three scalar fields. Using this simple supergravity model we construct families of supersymmetric solutions that could serve as candidate descriptions of the circular D0-branes. Unfortunately, all these solutions are singular and their physical interpretation is unclear.\footnote{We were informed by Juan Maldacena and Jorge Santos that they have independently constructed some of these solutions in unpublished work.} To remedy this impasse we propose an analytic continuation of this 3-scalar gauged supergravity model to general values of the dimension $d$. This formal analytic continuation is implemented directly at the level of the supergravity BPS equations that can be solved numerically for general values of $d$ and leads to cigar-like gravitational backgrounds that are smooth in the interior and should be viewed as the supergravity dual to the MSYM theory on $S^d$. This supergravity ``dimensional regularisation'' proves very useful since it regulates the singularity encountered for $d=1$ and allows for the holographic calculation of physical observables. To this end we follow \cite{Bobev:2018ugk,Bobev:2019bvq} and implement the holographic renormalisation procedure for non-conformal branes introduced in \cite{Kanitscheider:2008kd} to calculate the supergravity on-shell action and the regularised action of a fundamental string wrapping the equator of the boundary $S^d$. These two calculations are in excellent agreement with the QFT supersymmetric localisation results on $S^d$ and in the $d \to 1$ limit reproduce the finite result in~\eqref{eq:FW3-prediction}.

Our setup appears to be related to the BMN matrix quantum mechanics but we fail to establish a very precise connection to this well-studied model. For instance, we show that the 11d uplift of our 2d singular supergravity solutions is compatible with the general form of supersymmetric solutions with 16 supercharges and $\su(1,1)\times \so(6)$ global symmetry discussed in \cite{Lin:2005nh}, but since the solution is singular, its 11d interpretation is unclear. Supersymmetric localisation has been applied to the BMN matrix quantum mechanics in \cite{Asano:2012zt,Asano:2014vba,Asano:2014eca} but the resulting path integral was not analysed at large $N$ and it is not clear how to relate these results to our matrix model arising via analytic continuation in dimension. Recently the supersymmetric index of the BMN quantum mechanics was calculated in \cite{Chang:2024lkw} and since this is an $S^1$ partition function one may expect a relation to our results. The logarithm of the index in \cite{Chang:2024lkw} appears to scale differently with $N$ in the large $N$ limit when compared to the $N^2$ scaling of the free energy in \eqref{eq:FlogWintro} and suggests that either the two quantities are distinct, or they are related by an additional Casimir energy type prefactor not included in the analysis of \cite{Chang:2024lkw}. We also note that the BMN matrix quantum mechanics on $S^1$ has been studied extensively with various approaches ranging from lattice QFT to holography and supergravity, see \cite{Kimura:2003um,Anagnostopoulos:2007fw,Catterall:2007fp,Catterall:2008yz,Hanada:2008ez,Catterall:2009xn,Hanada:2011fq,Hanada:2013rga,Costa:2014wya,Berkowitz:2018qhn,Pateloudis:2022oos,Komatsu:2024vnb,Dias:2024vsc} for a selection of relevant references. These studies however focus on the thermal physics of the model which should be different from the supersymmetric setup on $S^1$ that we consider here.

We start our presentation in the next section with a review on the MSYM theory on $S^d$ and the matrix model obtained by supersymmetric localisation. We also discuss the analytic continuation in dimension and the relation to the BMN matrix quantum mechanics in the $d \to 1$ limit. In Section~\ref{sec:dual} we present the supergravity solutions dual to the MSYM theory on $S^d$ and discuss the singularity arising in the $d \to 1$ limit. The holographic analysis of these supergravity solutions is presented in Section~\ref{sec:FandWL} where we show that the holographic results are in agreement with the localisation calculations. In Section~\ref{sec:thermal} we review the thermal D-brane supergravity solutions and study them from the perspective of analytic continuation in dimensions. We conclude with a short discussion and outlook in Section~\ref{sec:conclusion}. Various technical details of the relevant supergravity theories and the spherical brane solutions are relegated to the two appendices.

%%%%%%%%%%%%%%%%%%%%%%%%%%%%%%%%%
\section{MSYM on \texorpdfstring{$S^d$}{Sd} and the BMN matrix quantum mechanics}          
\label{sec:BMN}	    
%%%%%%%%%%%%%%%%%%%%%%%%%%%%%%%%%

In the seminal paper \cite{Blau:2000xg}, Blau constructed the Lagrangian of supersymmetric Yang-Mills theory (SYM) in any dimension $d\le7$ on curved spaces. In particular, this construction, and more precisely the Family B in \cite{Blau:2000xg}, can be applied to study the Euclidean MSYM on the round sphere $S^d$ or hyperbolic space $\mathbf{H}^d$. In this section we will briefly review Blau's construction and the vacuum structure of the corresponding MSYM theories. We then discuss how to apply supersymmetric localisation to MSYM on $S^d$ and then turn to a discussion to the relation of this construction in the $d \to 1$ limit to the mass deformed BFSS model \cite{Banks:1996vh}, i.e. the BMN matrix quantum mechanics \cite{Berenstein:2002jq}.

%%%%%%%%%%%
\subsection{Euclidean MSYM on a sphere}
%%%%%%%%%%%

The simplest way to obtain the action of the Euclidean maximally supersymmetric Yang-Mills theory in flat space is to start with the ${\cal N}=1$ SYM in ten dimensions and consider its dimensional reduction. In ten dimensions the vector multiplet consists of a vector $A_M$ with $M=0,1,\ldots, 9$ and a Majorana-Weyl fermion $\Psi$.\footnote{We will suppress all spinor indices in the formulae below.} We are interested in a supersymmetric non-Abelian gauge theory and thus both of these fields transform in the adjoint of the gauge group which for concreteness we consider to be $\SU(N)$. The Lagrangian of the ten-dimensional theory on $\mathbf{R}^{1,9}$ is
\begin{equation}\label{eq:MSYM-Lag-flatspace}
\cL_\text{SYM} = -\f{1}{g_\text{YM}^2} \Tr\bigg( \f12 F_{MN}F^{MN} - \bar\Psi \slashed D \Psi\bigg)\,,
\end{equation}
where $g_\text{YM}$ is the gauge coupling and we use Hermitian generators of the gauge group such that $F_{MN} = \partial_M A_N-\partial_N A_M -\ii [A_M,A_N]$ and $D_M\Psi = \partial_M \Psi -\ii [A_M,\Psi]$. 
The Lagrangian \eqref{eq:MSYM-Lag-flatspace} is invariant under the following supersymmetry transformations
\begin{equation}\label{eq:MSYM-susy-flatspace}
\delta A_M= \bar\epsilon\Gamma_M\Psi - \bar{\Psi}\Gamma_M\epsilon\,,\qquad \delta \Psi=\slashed F\epsilon\,,
\end{equation}
where $\epsilon$ is a constant Majorana-Weyl supersymmetry transformation parameter. In ten dimensions, Majorana-Weyl spinors have 16 independent components and so the ${\cal N}=1$ SYM theory in ten dimensions is maximally supersymmetric.

The dimensional reduction of the ten-dimensional theory to $d<10$ dimensions gives rise to an MSYM theory in lower dimensions. The Lagrangian is obtained by simply assuming that the fields do not depend on $10-d$ of the coordinates. To write the Lagrangian in $d$ dimensions we split up the gauge field into $A_M = (A_\mu,\phi_I)$ where $A_\mu$ is interpreted as a $d$-dimensional gauge field and $\phi_I$ are $10-d$ scalar fields. Likewise, we split up the 10d fermion into multiple lower-dimensional fermions but we will not present the details of this procedure since it depends on the value of $d$. With this at hand, we can expand the Lagrangian \eqref{eq:MSYM-Lag-flatspace} in terms of the new fields and obtain the MSYM theory in $d$ dimensions. From now on we will keep this dimensional reduction procedure in mind but will continue using \eqref{eq:MSYM-Lag-flatspace} as a proxy for the lower-dimensional MSYM Lagrangian in flat space.

In addition to the Poincar\'e invariance in $d$-dimensions, the $d$-dimensional MSYM Lagrangian preserves supersymmetry and $\so(10-d)$ R-symmetry\footnote{The $d$-dimensional supersymmetry transformation can be derived by performing the dimensional reduction of the 10d transformations in \eqref{eq:MSYM-susy-flatspace}.} under which the scalars transform in the fundamental representation and the fermions transform according to the branching of the ${\bf 16}$ representation of $\so(1,9)$ into $\so(1,d-1)\times \so(10-d)$. An important subtlety arises if one wishes to consider the $d$-dimensional MSYM theory in Euclidean signature, which can be understood from its 10d avatar. In order to obtain the Euclidean theory in $d<10$ dimensions we must perform a formal reduction of the 10d theory along the time direction. This leads to the fact that the R-symmetry of the reduced theory is $\so(1,9-d)$ which is non-compact, and one of the scalars, which we will denote by $\phi_0$, has the wrong sign kinetic term. This seems to be an unavoidable feature of the $d$-dimensional Euclidean MSYM which can be traced to the fact that there is no Euclidean MSYM theory in ten (or higher) dimensions.\footnote{The minimal spinor in ten Euclidean dimensions has 32 real components and cannot be used to build a SYM theory without including gravity.} 

We now turn to the main focus of this paper which is Euclidean MSYM on the round sphere. As explained in \cite{Blau:2000xg} it is possible to write a modified SYM Lagrangian which preserves all 16 supercharges in $d\le7$ dimensions. The full Lagrangian reads\cite{Blau:2000xg}
\begin{equation}\label{eq:MSYM-Lag}
	\cL_{S^d} = \cL_\text{SYM} -\f{1}{g_\text{YM}^2\cR^2}\Tr\Big[ (d-2)\phi_I\phi^I + (d-4)\phi_a\phi^a\Big] + \f{(d-4)}{2g_\text{YM}^2\cR}\Tr\Big[ \bar\Psi\Gamma_{012}\Psi+8\ii\phi_0[\phi_1,\phi_2] \Big]\,.
\end{equation}
Here $\cL_\text{SYM}$ denotes the original MSYM Lagrangian \eqref{eq:MSYM-susy-flatspace} with the flat space metric replaced with the round metric on $S^d$ with radius $\mathcal{R}$. We use slightly different notation and conventions from the ones in \cite{Blau:2000xg}: we have reinstate the gauge coupling $g_\text{YM}$, use a different notation for the radius of $S^d$, namely $\alpha_{\rm there}=(2\cR)^{-1}$, and work with Hermitian generators of the gauge algebra as opposed to the anti-Hermitian convention in \cite{Blau:2000xg}. The latter introduces an additional factor of $\ii$ in the cubic scalar coupling, see for instance (3.20) of \cite{Blau:2000xg}. The index $I=0,1,\ldots, 9-d$ goes over all scalar fields, while the index $a=0,1,2$ singles out three of the scalars which acquire a non-conformal mass-term and a cubic coupling. The existence of this SYM Lagrangian only for $d\le 7$ can be traced to the fact that this cubic coupling requires three different scalars. In addition, in order to preserve supersymmetry the fermions acquire a mass term. It is important to note that these mass terms are not $\so(1,9-d)$ invariant but preserve only the $\su(1,1)\times \so(7-d)$ subalgebra.%
\footnote{\label{fn:su2vssu11}Note that the Lie algebras $\su(2)\sim \so(3)$ and $\su(1,1)\sim \so(1,2)$ are different real forms of the complex Lie algebra $\sl(2,\bC)$. Let the generators of $\sl(2,\bC)$ be $\mathfrak{t}_a$ with $a=1,2,3$, then the $\sl(2,\bC)$  Lie bracket is given by
	\[
		\comm{\mathfrak{t}_a}{\mathfrak{t}_b} = \ii \epsilon_{abc}\,\mathfrak{t}^c\,.
	\]
	When choosing a real forms $\su(2)$ or $\su(1,1)$, the commutation relation above are unchanged. The only difference is that for $\su(1,1)$ the indices are raised with the Killing form $g^{ab}=\diag(-1,1,1)$ while for $\su(2)$ one should use the identity matrix.
} 
Finally, note that when the scalar indices $a$ or $I$ are raised this has to be done with the Minkowski metric which in our convention has a $-1$ in the $00$ component. Importantly, the value $d=4$ is special since the cubic coupling and the non-conformal scalar mass term vanish. Indeed, for $d=4$ the MSYM theory is conformal and since $S^4$ is conformally flat one simply needs to add the $\phi_I\phi^I$ conformal mass term for the six scalars in order to preserve full superconformal invariance.

A closely related MSYM theory to the one presented above can be defined on Riemannian manifolds of constant \emph{negative} curvature, i.e. ${\bf H}^d$, see (3.20) of \cite{Blau:2000xg}. In this case the MSYM Lagrangian takes the form
\begin{equation}
	\label{eq:mass-Hyperbolic}
	\cL_{{\bf H}^d} = \cL_\text{SYM}+\f{1}{g_\text{YM}^2\cR^2}\Tr\Big[ (d-2)\phi_I\phi^I + (d-4)\phi_a\phi^a\Big] - \f{(d-4)}{2g_\text{YM}^2\cR}\Tr\Big[ \bar\Psi\Gamma_{123}\Psi+8\ii\phi_1[\phi_2,\phi_3] \Big]\,,
\end{equation}
where the index $a$ runs over $1,2,3$ which does not include the scalar $\phi_0$ with negative kinetic term. The MSYM theories on $S^d$ and ${\bf H}^d$ have important subtle differences. Crucially, the R-symmetry is $\su(1,1)\times \so(7-d)$ for the former and $\su(2)\times \so(1,6-d)$ for the latter. Finally we remark that the Lorentzian MSYM theory can be placed on manifolds of constant negative curvature, i.e. AdS, preserving 16 supercharges. Formally the Lagrangian is identical to \eqref{eq:mass-Hyperbolic} except that the index $I$ only runs over spacelike scalars with positive sign kinetic term. This Lorentzian theory exists in $d\le 7$ dimensions and has R-symmetry $\su(2)\times \so(7-d)$, while the Euclidean theory on ${\bf H}^d$ exists only for $d\leq 6$.

%%%%%%%%%%%
\subsection{Supersymmetric vacua}\label{subsec:vacua}
%%%%%%%%%%%

We now turn to a discussion of the classical supersymmetric vacua of the MSYM theories presented above. We hasten to add that there is no clear notion of a Hamiltonian for an Euclidean  QFT on $S^d$ and thus no clear notion of a vacuum state. Nevertheless, we will refer to a supersymmetric solution of the classical equations of motion as a supersymmetric vacuum. To find such supersymmetric vacua we need to impose that the supersymmetry variation of the fermions vanish. The Lagrangian \eqref{eq:MSYM-Lag} is invariant under the supersymmetry transformation\cite{Blau:2000xg}
\begin{equation}
\delta A_M = \bar\epsilon\Gamma_M\Psi - \bar{\Psi}\Gamma_M\epsilon\,,\qquad \delta \Psi=\slashed F\epsilon+\f{2}{\cR}\Big(\phi_I\Gamma^I+(d-4)\phi_a\Gamma^a\Big)\Gamma_{012}\epsilon\,,
\end{equation}
where $\epsilon$ is a conformal Killing spinor on $S^d$. A supersymmetric field configuration satisfies $\delta\Psi=\Psi=0$ which implies
\begin{equation}\label{eq:Sdsusyvac}
A_\mu=0\,,\qquad \phi_{I\ne a}=0\,,\qquad [\phi_a,\phi_b] = \f{d-3}{\cR}\ii\epsilon_{abc}\phi^c\,.
\end{equation}
One can show that this supersymmetric field configuration also solves the equations of motion. If we rescale the scalars as
\begin{equation}\label{eq:tadef}
	\phi_a = \f{(d-3)}{\cR}t_a\,,
\end{equation}
we find that to solve the last equation in \eqref{eq:Sdsusyvac} $t_a$ must furnish a representation of $\su(1,1)$.\footnote{Clearly $d=3$ is a special case for which any set of commuting matrices for $\phi_a$ will lead to a supersymmetric vacuum solution.} This may naively suggest that we have a collection of non-trivial supersymmetric vacua. However, if the gauge group is $\SU(N)$ the scalars $\phi_a$ are in the adjoint representation and thus $t_a$ are Hermitian $N\times N$ matrices. This in turn implies that they cannot furnish a non-trivial representation of $\su(1,1)$. We are therefore led to conclude that for MSYM on $S^d$ the only supersymmetric vacuum is the trivial one, i.e. $\phi_a=0$. This is in line with the common expectation that supersymmetric QFTs on $S^d$ have a trivial vacuum moduli space since the curvature of $S^d$ lifts any vacuum degeneracy that may be present for the theory on $\mathbf{R}^d$.

The supersymmetric vacuum equations on ${\bf H}^d$ and AdS$_d$ are structurally very similar to the ones on $S^d$ with an important difference. Due to the different signs in the interaction terms in the Lagrangian \eqref{eq:mass-Hyperbolic} one finds that the scalars $\phi_{1,2,3}$ must furnish a representation of $\su(2)$, see \cite{Blau:2000xg}. This leads to a number of non-trivial vacua with non-vanishing scalar field vevs. This type of vacuum equation for adjoint scalars in supersymmetric QFT is a familiar predicament from the physics the 4d $\mathcal{N}=1^{*}$ SYM theory \cite{Donagi:1995cf,Polchinski:2000uf} and from the BMN matrix quantum mechanics \cite{Berenstein:2002jq,Lin:2005nh}. To be more explicit,
since the scalars $\phi_a$ transform in the adjoint representation of the gauge group which we take to be $\SU(N)$ the $t_a$ in \eqref{eq:tadef} must form an $N$-dimensional representation of $\su(2)$. Any such representation can be decomposed as a direct sum of irreducible representations, and is labelled by a partition of $N$ such that $N=\sum D\, k_D$, where $k_D$ counts the number of times the irreducible representation of dimension $D$ appears. In all but one of these vacua the scalars $\phi_a$ are non-zero and thus the gauge group is spontaneously broken. It will be very interesting to understand further the structure of these vacua for MSYM on ${\bf H}^d$ and their potential realisation as polarised ``hyperbolic'' D$p$-branes due to the Myers effect, see \cite{Myers:1999ps,Polchinski:2000uf}.

The discussion above is formally valid for general values of $d$ and points to a very drastic difference in the low-energy physics of the MSYM theory on $S^d$ and ${\bf H}^d$. This may seem somewhat surprising if we analytically continue to $d=1$ and especially to $d=0$ where these two manifolds are either flat or simply the same. We attribute this stark difference to the structure of the R-symmetry algebra of the theories on $S^d$ and ${\bf H}^d$ under which the scalars $\phi_a$ transform differently. From now on, when we analytically continue to $d=0$ or $d=1$ we will assume that we do so by starting with the theory on $S^d$ and thus we will have only a single supersymmetric vacuum with $\phi_a=0$.

%%%%%%%%%%%
\subsection{Supersymmetric localisation on the sphere}\label{subsec:susyloc}
%%%%%%%%%%%

In \cite{Minahan:2015jta,Minahan:2015any} it was shown that supersymmetric localisation can be used to reduce the path integral of the MSYM theory on $S^d$ defined by the Lagrangian in  \eqref{eq:MSYM-Lag} to a Hermitian matrix model. Introducing the dimensionless matrix $\sigma = \cR \phi_0$ and the dimensionless 't~Hooft coupling
\begin{equation}\label{eq:lambdaQFTdef}
\lambda = g_\text{YM}^2 N \cR^{4-d}\,,
\end{equation}
the partition function reduces to its saddle point expansion around the localising locus:
\begin{equation}\label{eq:Zsusylocdef}
Z = \int [\dd\sigma]\exp\Big( -\f{4\pi^2 N \vol_{d-4}}{\lambda}\Tr\sigma^2 \Big) Z_\text{1--loop}Z_\text{instantons}\,,
\end{equation}
where the integral ranges over the eigenvalues of $\sigma$. Here we have defined the volume of the unit $d$-sphere, 
\begin{equation}\label{eq:defVd}
\vol_{d}\equiv \frac{2\pi^{(d+1)/2}}{\Gamma\left(\f{d+1}{2}\right)}\,.
\end{equation}
$Z_\text{1--loop}$ gives the contribution of one-loop fluctuations around the localising locus and $Z_\text{instantons}$ gives the contributions of subleading instanton saddles. In the large $N$ limit, the contribution of instantons can be ignored and the partition function is dominated by the leading supersymmetric locus which has $A_\mu=0$. Despite this simplification, we are still left with a complicated integral over the matrix $\sigma$. The large $N$ limit allows to perform yet another saddle point expansion, where the dominant contribution to the integral can be obtained by solving
\begin{equation}\label{eq:densityeq}
4\pi^{2}N \vol_{d-4}\sigma_i = \sum_{j\ne i} G(\sigma_i-\sigma_j)\,,
\end{equation}
with $\sigma_i$ the eigenvalues of the $N\times N$ matrix $\sigma$ and the kernel $G$ takes the form \cite{Minahan:2015any}
\begin{equation}\label{eq:Kerndef}
\f{\ii G(\sigma)}{\Gamma(4-d)} = \f{\Gamma(d-3+\ii\sigma)}{\Gamma(1+\ii\sigma)}-\f{\Gamma(d-3-\ii\sigma)}{\Gamma(1-\ii\sigma)}- \f{\Gamma(\ii\sigma)}{\Gamma(4-d+\ii\sigma)}+ \f{\Gamma(-\ii\sigma)}{\Gamma(4-d-\ii\sigma)}\,.
\end{equation}
The equation \eqref{eq:densityeq} is easier to analyse in terms of an almost continuous eigenvalue distribution $\rho$ defined as
\begin{equation}
\rho(\sigma) = \f{1}{N} \sum_{i}\delta(\sigma-\sigma_i)\,,
\end{equation}
which allows to transform the equation \eqref{eq:densityeq} into an integral equation for the eigenvalue distribution $\rho(\sigma)$.

In the strong coupling limit, $\lambda\gg1$, of interest for holographic applications, the eigenvalues $\sigma$ are well separated and one can approximate $G$ by its large argument expression. In this regime, and for $3<d<6$, the eigenvalue density can be expressed as \cite{Bobev:2019bvq} 
\begin{equation}
	\label{eq:rho-and-b}
\rho = \f{\vol_{d-2}}{\lambda \Gamma(6-d)}(b_d^2-\sigma^2)^{\f{5-d}{2}}\,,~~\qquad~~ \Big(\f{b_d}{2\pi}\Big)^{6-d} = \f{2\lambda}{\vol_{d-2}\vol_{5-d}\vol_{7-d}}\,,
\end{equation}
where we have used \eqref{eq:defVd} to write some of the $\Gamma$-functions in terms of sphere volumes. Equipped with this solution one can proceed to compute the sphere free energy of the MSYM theory as well as the vacuum expectation value of the $\f12$-BPS Wilson loop in the fundamental representation of the gauge group to leading order in $N$ and the large coupling $\lambda$
\begin{equation}\label{eq:genmatrixmodelprediction}
F_d = -\f{\pi N^2\vol_{d-2}}{\lambda}\f{(6-d)(3-d)}{(8-d)(4-d)}b_d^2\,,\qquad \log \langle W\rangle = 2\pi b_d\,.
\end{equation}
The scaling of these quantities with the 't~Hooft coupling $\lambda$ is $F_d \sim N^2 \lambda^{\f{d-4}{6-d}}$ and $\log \langle W\rangle \sim \lambda^{\f{1}{6-d}}$ which is in line with the supergravity scaling similarity arguments discussed in \cite{Biggs:2023sqw}. It should be noted that the general analysis presented here is only valid for $3<d<6$. In \cite{Bobev:2019bvq}, further analysis was performed for the integer dimensions $d=3,6,7$ and it was demonstrated that \eqref{eq:genmatrixmodelprediction} also holds in these cases.\footnote{We remark that the holographic limit for $d=7$ is not at large $\lambda$ and one has to renormalise the coupling appropriately. We refer to \cite{Bobev:2019bvq,Minahan:2022rsx} for more details.} The result in \eqref{eq:genmatrixmodelprediction} was further corroborated by deriving it using holography and the spherical brane solutions in supergravity \cite{Bobev:2018ugk,Bobev:2019bvq}.  

The supersymmetric localisation results in \eqref{eq:genmatrixmodelprediction} are derived rigorously in integer dimensions. Given their final form however, it is very tempting to treat the dimension $d$ as a real parameter and consider \eqref{eq:genmatrixmodelprediction} to be valid for general values of $d$. Indeed, as we will show below, holography supports this idea and allows for the derivation of \eqref{eq:genmatrixmodelprediction} by using an analytic continuation of a simple supergravity model for the range $1<d<4$.

In principle the matrix model \eqref{eq:Zsusylocdef} can be used to study subleading corrections to the large $N$ and large $\lambda$ expressions in \eqref{eq:genmatrixmodelprediction}. It turns out that the subleading corrections in powers of $\lambda$ are calculationally more accessible. They can be obtained by solving the saddle point equation \eqref{eq:densityeq} for the eigenvalue density $\rho$ either exactly or in a series expansion for large $\lambda$. Exact solutions exist for $d=3,4,5$ dimensions, see \cite{Bobev:2019bvq,Gautason:2021vfc} for a discussion, and some recent progress has been made for $d=7$ \cite{Astesiano:2024sgi}. It would be very interesting to understand further these perturbative corrections or even the full analytic solution for general $d$. Using holography, these results will shed light on the structure of the $\alpha'$ and $g_{s}$  corrections to string theory in certain RR backgrounds. 

%%%%%%%%%%%
\subsection{Relation to the BMN matrix quantum mechanics}
%%%%%%%%%%%

The BMN matrix quantum mechanics is a Lorentzian theory obtained by adding an $\su(2)\times \so(6)$ invariant set of interaction terms to the BFSS model while preserving all 16 supercharges \cite{Berenstein:2002jq}. Interestingly, if we analytically continue the MSYM Lagrangian on AdS$_d$ in \eqref{eq:mass-Hyperbolic} to $d=1$ we obtain precisely the Lagrangian of the BMN model where the BMN mass $\mu$ and the AdS scale $\mathcal{R}$ are related as, see Equation (5.2) in \cite{Berenstein:2002jq},
\begin{equation}
	\mu = \f{6}{\cR}\,.    
\end{equation}
As emphasised in Section~\ref{subsec:vacua} the BMN model has a rich set of supersymmetric vacua corresponding to various polarised brane that have a precise supergravity description as explained in \cite{Lin:2005nh}. Here, we are however interested in taking the $d \to 1$ limit of the MSYM theory on $S^d$. This leads to a Euclidean Lagrangian that is closely related but distinct from the BMN one, see \eqref{eq:MSYM-Lag}. In particular, the R-symmetry is $\su(1,1)\times \so(6)$ and there are no non-trivial supersymmetric vacua if the gauge group is $\SU(N)$. We will refer to this model as Euclidean BMN quantum mechanics on $S^1$.

The $S^d$ supersymmetric localisation results in Section~\ref{subsec:susyloc} can be analytically continued to $d=1$ to find some of the physical observables in the Euclidean BMN quantum mechanics. Taking the $d \to 1$ limit in the matrix model defined by \eqref{eq:Zsusylocdef} and \eqref{eq:Kerndef} is somewhat subtle and should be done carefully since various quantities vanish or diverge in this limit. The saddle point equation \eqref{eq:densityeq} for the eigenvalues $\sigma_i$ takes the form
\begin{equation}\label{eq:d1matrixmodelsaddle}
N \sigma_i = \lambda \sum_{j\ne i} G(\sigma_i-\sigma_j)\,,\qquad G(\sigma)  = -\f{3\coth \pi\sigma}{(\sigma^2+4)(\sigma^2+1)}\,,
\end{equation}
where we have incorporated the numerical factors on the left-hand side of \eqref{eq:densityeq} into the kernel $G(\sigma)$ to obtain a regular $d \to 1$ limit. At weak coupling, eigenvalue separation should be small, and the kernel can be approximated by its small $\sigma$ expansion $G(\sigma)\to -3/(4\pi\sigma)$. In this limit the eigenvalue density is given by the Wigner semi-circle distribution. We are more interested in the large $\sigma$ expansion of $G$ which corresponds to the strong coupling behaviour of the theory. In this limit we find $G(\sigma)\to -3\,\sgn(\sigma)/\sigma^4$. It is not difficult to see that this expression for the kernel implies that there are no real solutions to \eqref{eq:d1matrixmodelsaddle} at strong coupling. There are however complex solutions with an intricate structure. Analysing these complex saddles in detail is beyond the scope of the current work. Instead, we proceed by analytically continuing the general $d$ results in \eqref{eq:genmatrixmodelprediction} to $d=1$. We take $d=1+\epsilon$ with $\epsilon \to 0$ and find that the free energy and Wilson loop vev in \eqref{eq:genmatrixmodelprediction} can be written as
\begin{equation}
	\label{eq:FlogW-Sd}
		F_{1+\epsilon} = -\f{5\pi N^2}{7}\bigg(\f{200\epsilon^3}{3\lambda^3}\bigg)^{1/5}\,,\qquad
		\log \vev{W_{1+\epsilon}} = \pi \bigg( \f{720\lambda}{\epsilon} \bigg)^{1/5}\,.
\end{equation}
To make these expressions regular in the $\epsilon \to 0$ limit one can change the 't Hooft coupling into a new ``regularised'' coupling $\tilde{\lambda}= \lambda/\epsilon$. This is clearly a somewhat arbitrary ``regularisation scheme'' which we cannot properly justify. Indeed, as we discuss in detail below, the holographic dual spherical brane supergravity solutions with $S^d$ worldvolume also exhibit singularities when analytically continued to $d=1$. In the absence of a better justification of this regularisation procedure we can instead focus on the scheme independent quantity $F (\log \vev W)^3$ which is given by
\begin{equation}\label{eq:FlogWres}
	F_{1+\epsilon} (\log \vev{W_{1+\epsilon}})^3 = -\f{600\pi^{4}N^2}{7}\,.
\end{equation}
This is a finite quantity that should be physical and scheme independent. Indeed, as we show in Section~\ref{sec:FandWL}, the result in \eqref{eq:FlogWres} can be derived using a holographic calculation. To gain some confidence in the results in \eqref{eq:FlogW-Sd} we note that the scaling of both the free energy and Wilson loop vev with the strong coupling $\lambda$ are the ones expected from the scaling similarity exhibited by the dual type IIA supergravity, see \cite{Biggs:2023sqw}.

We should note that there are studies of the BMN matrix quantum mechanics via supersymmetric localisation, see \cite{Asano:2012zt,Asano:2014eca}, as well as calculation of the supersymmetric index of the model on $S^1$, see \cite{Chang:2024lkw} for a recent account. One could perhaps expect that there is a relation between these calculations and the analysis we presented above. Unfortunately, it is not clear to us how to relate these disparate analyses. Indeed, the results we find for the $S^1$ free energy of our Euclidean BMN model appear to be quite different from the ones for the supersymmetric index of the theory as presented in \cite{Chang:2024lkw}. Since the index is independent of the coupling, it can be computed at weak coupling where the Lorentzian BMN matrix model splits in different superselection sectors labelled by the supersymmetric vacua. The full BMN index is then given by
\begin{equation}
	\cI = \sum_{i\in\text{vacua}} \cI_i\,.
\end{equation}
Since the BMN matrix quantum mechanics is gapped, there is a unique ground state in each superselection sector such that in the unrefined limit of the index, i.e. when all R-symmetry fugacities are switched off, we have $\cI_i=1$ and $\cI \simeq p(N) \sim \exp \left(\sqrt N \right)$, where $p(N)$ is the function counting the number of partitions of $N$ which at large $N$ can be approximated by the asymptotic formula of Hardy-Ramanujan. Clearly, this expression is not compatible with the $\exp N^2$ growth exhibited by the $S^1$ free energy computed in \eqref{eq:FlogW-Sd}. This observation suggests that there is a crucial difference between the index in \cite{Chang:2024lkw} and the $S^1$ partition function computed by our analytically continued MSYM $S^d$ matrix model. Perhaps the discrepancy in the large $N$ scaling of the two observables can be attributed to a supersymmetric Casimir energy prefactor. It is also possible that the complex saddles of the MSYM $S^d$ matrix model in the $d \to 1$ limit discussed above \eqref{eq:FlogW-Sd} play an important role when one takes the large $N$ limit.

We end this discussion with some brief comments on the potential relevance of our results to the IKKT matrix model \cite{Ishibashi:1996xs}. This matrix model preserves maximal supersymmetry in $d=0$ and  thus should be somehow related to the $d \to 0$ limit of the MSYM theory on $S^d$. Indeed, taking $d=0$ in the MSYM Lagrangian in \eqref{eq:MSYM-Lag} one finds a matrix model with 16 supercharges and $\su(1,1)\times \so(7)$ R-symmetry. This is not precisely the IKKT theory but rather, a BMN-like mass deformation of it. This type of deformations of the $\so(10)$ invariant IKKT model have not been extensively studied in the literature, see however \cite{Bonelli:2002tb,Kumar:2022giw} and especially the recent work in \cite{Hartnoll:2024csr,Komatsu:2024bop}, and it will be interesting to understand these models better. Here we simply note that the $d \to 0$ limit of the free energy in \eqref{eq:FlogW-Sd} is regular and one finds
\begin{equation}
	F_0 = -\f{3^{7/3}}{2^{5/3}}\f{N^2}{(-\lambda)^{2/3}}\,.
\end{equation}
This result has the correct scaling similar behaviour in the coupling $\lambda$ as dictated by the analysis in \cite{Biggs:2023sqw} and in Section~\ref{sec:FandWL} we discuss how it can potentially be derived holographically from the analytically continued spherical brane solutions.

%%%%%%%%%%%%%%%%%%%%%%%%%%%%%%%%%
\section{The supergravity dual}          
\label{sec:dual}	    
%%%%%%%%%%%%%%%%%%%%%%%%%%%%%%%%%

Given the setup described above, the natural place to look for a holographic dual to the Euclidean BMN matrix quantum mechanics is the $\su(1,1)\times \so(6)$ invariant truncation of the maximal $\so(1,8)$ gauged two-dimensional supergravity~\cite{Ortiz:2012ib}. This truncation consist of the metric and three scalars, a dilaton $\rho$ and two real scalars $X$ and $Y$ coming from the $\su(1,1)\times \so(6)$ singlets in the $\bf{44}$ and $\bf{84}$ of $\so(1,8)$ respectively. For a detailed description of this truncation we refer the reader to Appendix \ref{app:2dSUGRA}. The Euclidean action for this two-dimensional model can be written as\footnote{For future purposes this form of the action will be most useful. However, it is more common in two-dimensional dilaton-gravity theories to remove the dilaton kinetic term which can be done through a simple redefinition of the scalar fields, see Appendix~\ref{app:2dSUGRA}.}
\begin{equation}
	\label{eq:action-0}
	S_{\rm 2d} = \f{1}{2\kappa_2^2} \int \dvol_2\, \rho \left( R-\f12 P_\mu P^\mu - V \right) + S_{\rm GH}\,,\qquad S_{\rm GH} = \f{1}{\kappa_2^2}\int \dvol_1 \rho K\,.
\end{equation}
where $\dvol_n$ is the $n$-dimensional volume form and we have added the Gibbons-Hawking term $S_{\rm GH}$ which is needed to study solutions with asymptotic boundaries. We have also defined $P_{\mu}$ as
\begin{equation}
	P_\mu = \f{\partial_\mu X-\partial_\mu Y}{X} - \f{Y\partial_\mu \rho}{3X\rho}\,,
\end{equation}
and the potential $V$ can be written in terms of real superpotentials $W$ and $\wti W$ as explained in Appendix \ref{app:2dSUGRA} and takes the explicit form
\begin{align}
	W =&\, g\f{2+X-Y}{X^{1/3}\rho^{2/9}} \,,\qquad \wti{W} =\, g\f{2+X+Y}{X^{1/3}\rho^{2/9}} \,.\\	
	V =&\, -\f{3g^2}{2 X^{2/3}\rho^{4/9}}\big(8+12 X + X^2 -Y^2\big)\,,
	\end{align}
In Lorentzian signature $W$ and $\widetilde W$ would be related by complex conjugation and $g$ is the 2d gauged supergravity coupling constant.%
\footnote{The 2d gauged supergravity coupling can be related to the type IIA string theory parameters as 
	\begin{equation}
		(2\pi \ell_s g)^{-7} = \f{15N}{32\pi^4}\,.
	\end{equation}
} 
Let us emphasise that the $\so(1,8)$ invariant model, obtained by setting $X=1$, $Y=0$ is a model of 2d dilaton-gravity different from the JT gravity. Notably, there is no $\AdS_2$ vacuum. Instead the vacuum solution of this model has a running dilaton and corresponds to the dimensional reduction of the flat D0-brane solution of type IIA supergravity to two dimensions, see \eqref{eq:BPS2dXY0} and \eqref{eq:D0solAppA}. In particular, this solution provides a dual geometry suitable to study the BFSS matrix quantum mechanics. Turning on $X$ and $Y$ breaks the $\so(1,8)$ symmetry to $\su(1,1)\times \so(6)$ and can be understood as corresponding to the mass deformation leading to the BMN matrix quantum mechanics. More precisely, the three supergravity scalars can be understood as follows: the dilaton $\rho$ is dual to the gauge coupling constant in the matrix quantum mechanics, while the scalars $X$ and $Y$ are dual to the operators \cite{Ortiz:2014aja},
\begin{equation}\label{eq:dual-ops}
	\begin{aligned}
		\cO_X \propto&\,\f1N\Tr\left( \phi^a\phi_a - \f13  \phi^I\phi_I \right)\,,\\
		\cO_Y \propto&\, \f1 N \left[ \bar\Psi\Gamma_{012}\Psi+ \ii \Tr \phi_0 \comm{\phi_1}{\phi_2}\right]\,.  
	\end{aligned}
\end{equation}
Turning off the scalar $Y$ leads to a class of ``Coulomb branch'' supergravity solutions relevant for the BFSS matrix model, see \cite{Ortiz:2014aja} and Appendix~\ref{app:2dSUGRA} for more details. In order to make contact with the BMN matrix quantum mechanics it is therefore necessary to have non-trivial profiles for all three scalars. To construct such a solution, let us consider the following ansatz for the metric,
\begin{equation}
	\ds^2 = \dd r^2 + \cR^2\e^{2A(r)}\dd \tau^2\,,
\end{equation}
where $\tau \sim \tau + 2\pi$ is a coordinate on the circle with radius $\cR$ and the metric function $A(r)$ only depends on the radial coordinate $r$. To find the solutions of interest it is convenient to use the scalar $X$ as the radial coordinate, in terms of which the supergravity BPS equations can be rewritten as
\begin{equation}
	\label{eq:BPS0}
	\begin{aligned}
		\f{\dd Y}{\dd X} =&\, -\f{Y}{2X}\f{4+4X+7X^2-Y^2}{4-2X-2X^2-Y^2} \,, \\
		\f{\dd \rho}{\dd X} =&\, \f{3\rho}{2X}\f{(2+X)^2-Y^2}{4-2X-2X^2-Y^2}\,,  \\
		\f{\dd A}{\dd X} =&\, \f{1}{6X}\f{7(2+X)^2-4Y^2}{4-2X-2X^2-Y^2} \,,\\
		\f{\dd X}{\dd r} =&\, \f{X^{2/3} g\left(2(-2+X+X^2)+Y^2\right)}{\rho^{2/9}\sqrt{(2+X)^2-Y^2}}  \,.
	\end{aligned}	
\end{equation}
As detailed in Appendix \ref{app:2dSUGRA}, assuming $Y(r)\neq 0$, we can solve the BPS equation for the metric function and find the explicit expression,
\begin{equation}
	\cR^2\e^{2A} = \f{X^{2/3}\rho^{4/9}}{g^2} \f{(2+X)^2-Y^2}{Y^2}\,,
\end{equation}
such that the 2d metric takes the form
\begin{equation}
	\ds_{2}^2 = \f{\rho^{4/9}((2+X)^2-Y^2)}{X^{4/3} g^2} \left( \f{\dd X^2}{\left(2(-2+X+X^2)+Y^2\right)^2} + \f{X^2}{Y^2}\dd \tau^2 \right)\,. 
\end{equation}
The system of BPS equations has therefore been reduced to a single differential equation for $Y$ as a function of $X$. Once $Y$ is known, the dilaton can be obtained by simple integration. Before attempting to solve the BPS equation for $Y$, let us first study the critical points of the system. These are identified as the locations where both $X$ and $Y$ have fixed points. One way to identify such points is to demand that both the numerator and denominator of the right hand side of equations the first equation in \eqref{eq:BPS0} vanishes. There are five such points in the $(X,Y)$ plane,%
\footnote{All equations are symmetric under the interchange $Y\leftrightarrow -Y$. Here we discuss only the critical points with non-negative $Y$ but clearly there are analogous points with negative $Y$.}
\begin{equation}\label{eq:5critpts}
	(X,Y)= \left\{ (1,0)\,, (0,2)\,, \left( - \f23, \f{2\sqrt{10}}{3}\right)\,, (-2,0)\,, (0,0) \right\}\,.
\end{equation}
To identify the correct UV and IR loci let us study these five critical points in some more detail.

\begin{figure}[!htb]
	\centering
	\begin{tikzpicture}
		\node at (0,0) {\includegraphics[width=0.6\textwidth]{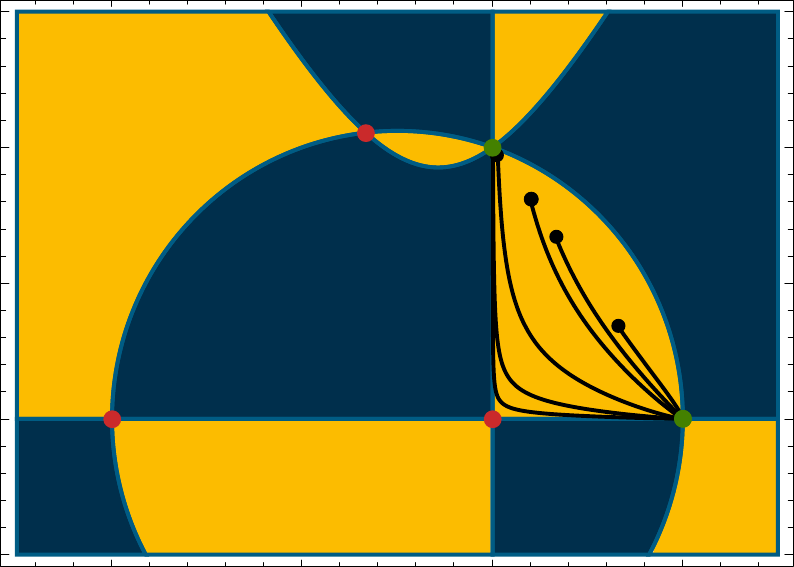}};
		\node at (4,-2) {\Large \textbf{UV}};
		\node at (1.5,2.2) {\Large \textbf{IR}};
		\node at (0,-4) {\Large $X$};
		\node at (-5.3,1) {\Large $Y$};
		\node[anchor=east] at (-4.8,3.3) {$3$};
		\node[anchor=east] at (-4.8,1.65) {$2$};
		\node[anchor=east] at (-4.8,0) {$1$};
		\node[anchor=east] at (-4.8,-1.65) {$0$};
		\node[anchor=east] at (-4.8,-3.3) {$-1$};
		\node[anchor=north] at (-3.55,-3.5) {$-2$};
		\node[anchor=north] at (-1.25,-3.5) {$-1$};
		\node[anchor=north] at (1.2,-3.5) {$0$};
		\node[anchor=north] at (3.45,-3.5) {$1$};
	\end{tikzpicture}
	\caption{A region plot showing the sign of the right-hand side of the BPS equation for $Y$ for $d=1$. Blue regions denote negative values of the right-hand side of the first equation in \eqref{eq:BPS0} while orange regions denote positive values. The five critical points in \eqref{eq:5critpts} are denoted with coloured dots located at the intersection of the solid lines separating different regions. The three unphysical fixed points are denoted with red dots, while the green dots represent the UV and IR loci at $(X,Y)_{\rm UV}=(1,0)$ and IR $(X,Y)_{\rm IR}=(0,2)$. The black curves represent the numerical solutions for $Y(X)$ using the equation \eqref{eq:BPS-p} for $d=\{3,2,1.6,1.08,1.007,1.0006\}$. As $d \to 1$ we see that the IR point indicated with the black solid dot approaches $(X,Y)_{\rm IR}=(0,2)$.}
	\label{fig:BPSregions}
\end{figure}
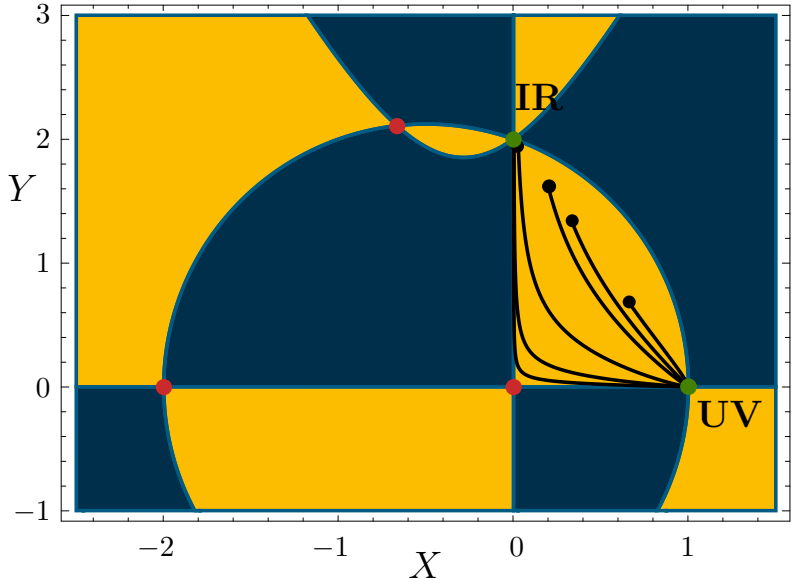

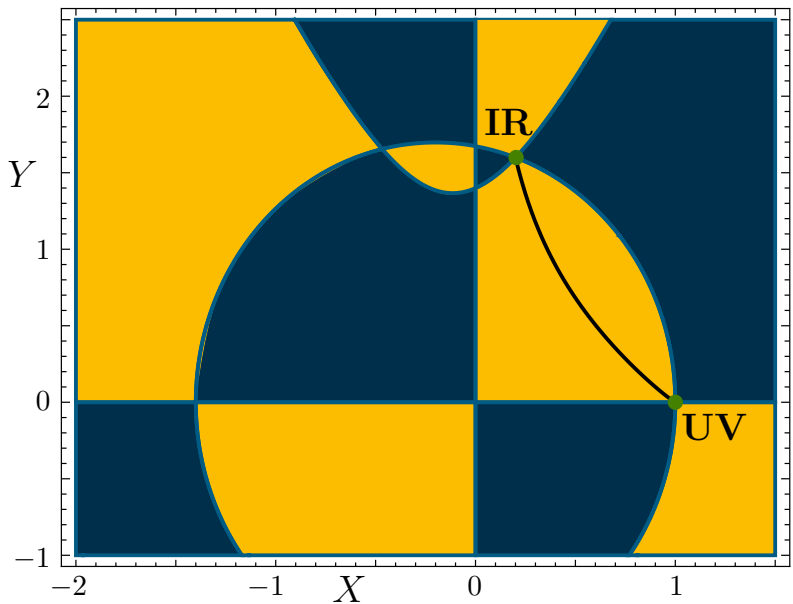
\begin{figure}[!htb]
	\centering
	\begin{tikzpicture}
		\node at (0,0) {\includegraphics[width=0.6\textwidth]{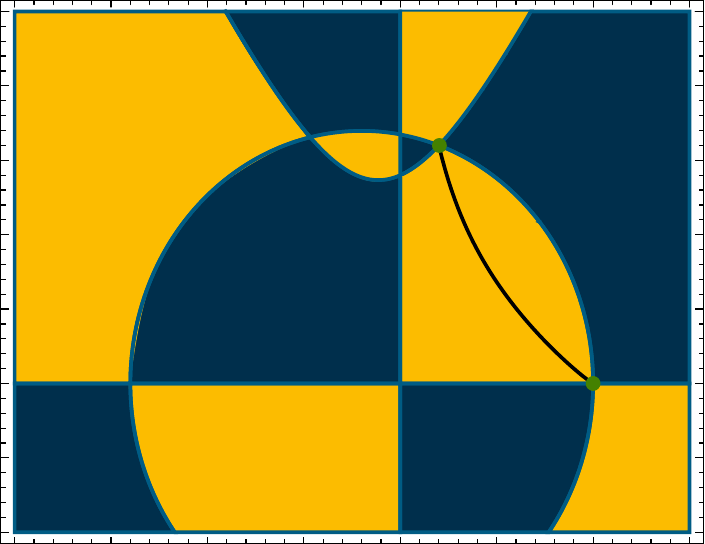}};
		\node at (3.8,-1.85) {\Large \textbf{UV}};
		\node at (1.1,2.2) {\Large \textbf{IR}};
		\node at (-1,-4) {\Large $X$};
		\node at (-5.3,1.5) {\Large $Y$};
		\node[anchor=east] at (-4.8,2.5) {$2$};
		\node[anchor=east] at (-4.8,0.5) {$1$};
		\node[anchor=east] at (-4.8,-1.5) {$0$};
		\node[anchor=east] at (-4.8,-3.6) {$-1$};
		\node[anchor=north] at (-4.7,-3.7) {$-2$};
		\node[anchor=north] at (-2.1,-3.7) {$-1$};
		\node[anchor=north] at (0.65,-3.7) {$0$};
		\node[anchor=north] at (3.3,-3.7) {$1$};
	\end{tikzpicture}
	\caption{A region plot showing the sign of the right-hand side of the BPS equation for $Y$ for $d=1.6$. Blue regions denote negative values of the right-hand side of the first equation in \eqref{eq:BPS-p} while orange regions denote positive values. The critical points are located at the intersection of lines separating different regions. The green dots represent the UV $(X,Y)_{\rm UV}=(1,0)$ and IR $(X,Y)_{\rm IR}=(0.2,1.6)$ loci, while the black curve is the numerical solution of the BPS equation  for $Y(X)$ that connects them.}
	\label{fig:BPSregiond16}
\end{figure}

\paragraph{UV region}
The point $(X,Y)_{\rm UV} = (1,0)$ corresponds to the UV region. Near this point both scalars become trivial and the solution reduces to the $\so(1,8)$ vacuum solution of the 2d gauged supergravity which is characterised by a simple profile for the dilaton as a function of the radial coordinate. The 10d uplift of this 2d supergravity solution corresponds to the type IIA supergravity solution describing the near horizon limit of coincident D0-branes, see \eqref{eq:BPS2dXY0} and \eqref{eq:D0solAppA}. Away from this UV point the scalars $X$ and $Y$ develop a radial profile and break the $\so(1,8)$ symmetry to $\su(1,1)\times\so(6)$.

\paragraph{IR region}
Inspired by the spherical brane solutions of \cite{Bobev:2018ugk} we expect that in the IR region the metric caps off smoothly as $\ds_2^2 \propto \dd r^2 + r^2\dd\tau^2$ and the scalar fields approach a finite constant value. These conditions alone however do not determine unambiguously the IR point since expanding the BPS equations around the four putative IR points we find that the metric is singular or the scalar fields diverge near all four of them. One could exclude the points $(-2,0)$ and $\left( - \f23, \f{2\sqrt{10}}{3}\right)$ since the canonically normalised scalar field $x\sim \log X$ becomes complex near them. A more unambiguous and convincing approach to fix the correct IR locus is to use the observation that the BMN matrix quantum mechanics can be thought of as the $d\rightarrow 1$ limit of MSYM on $S^d$. In a way that we will make precise momentarily, we can think of the 2d gravity model in \eqref{eq:action-0} as an analytic continuation in dimension of a family of gauged supergravity theories that lead to the spherical brane solutions of \cite{Bobev:2018ugk}. Employing this analytic continuation one finds unambiguously that the correct IR point, i.e. the locus where the Euclidean geometry smoothly caps off, is at
\begin{equation}
	(X,Y)_{\rm IR} = \lim_{d\to 1} \left( \f{d-1}{3}, \f{2(4-d)}3  \right) = (0,2)\,. 
\end{equation}
Analysing the supergravity BPS equations near this point is particularly difficult due to the fact that it lies at the intersection of three ``critical lines'' in Figure~\ref{fig:BPSregions} along which the right-hand side of the first equation in \eqref{eq:BPS0} vanishes or diverges. Moreover, while the metric is smooth in the IR region, the dilaton scalar $\rho$ blows up which further complicates the analysis and interpretation of any supergravity solution that asymptotes to this point. As we show below, for $d>1$ this triple intersection splits into two distinct critical points which ultimately resolves this singular behaviour and allows for a proper analysis of the IR region.

%%%%%%%%%%%
\subsection{Spherical D0-brane background}
\label{subsec:circularD0}
%%%%%%%%%%%

To understand better how to construct the circular D0-brane solution of interest, let us consider the spherical brane solutions of \cite{Bobev:2018ugk}. As reviewed in Appendix~\ref{app:Spherical-p}, we can write the supergravity BPS equations that determine these solutions for general $d<4$ as
\begin{equation}
	\label{eq:BPS-p}
	\begin{aligned}
		\f{\dd Y}{\dd X} &= -\f{Y}{2X}\f{(d-3)^2+4(2-d)X+7X^2-Y^2}{2(1-X)(3-d+X)-Y^2}\,,\\
		\f{\dd \eta}{\dd X} &= -\f{7-d}{2(d-1) X}\f{(3-d+X)^2-Y^2}{2(1-X)(3-d+X)-Y^2}\,,
	\end{aligned}
\end{equation}
while for the metric warp factor, $\cA$, we find the algebraic expression
\begin{equation}
	\cR^2\e^{2\cA} = \e^{-\f{2(4-d)}{7-d}\eta}\,X\,\f{ (3-d+X)^2-Y^2}{g^2 Y^2}\,.
\end{equation}
Using this, the metric can be written in terms of $X$ as
\begin{equation}
	\ds_{d+1}^2 = \f{((3-d+X)^2-Y^2)}{X g^2 }\e^{\f{-2(4-d)}{7-d}\eta}\left( \f{\dd X^2}{(2(1-X)(3-d+X)-Y^2)^2} + \f{X^2}{Y^2}\dd\Omega_{d}^2\right)\,.
\end{equation}
The BPS equations \eqref{eq:BPS-p} do not have a good $d\to 1$ limit. To remedy this we introduce the following field redefinition,
\begin{equation}
	\label{eq:maptorho}
	\rho = \e^{-\f{d-1}2 \eta}\,.
\end{equation}
Using this we find the following BPS equation for the scalar $\rho$, 
\begin{equation}
	\label{eq:BPS-rho}
	\f{\dd \rho}{\dd X} = \f{(7-d)\rho}{4X}\f{(3-d+X)^2-Y^2}{2(1-X)(3-d+X)-Y^2}\,.
\end{equation}
This equation has a good $d\to1$ limit and, together with the BPS equation for $Y$, reduces precisely to the system of BPS equations \eqref{eq:BPS0} of the 2d supergravity discussed above. Similarly, the metric functions as well as the full 2d metric are related as follows,
\begin{equation}
	\ds^2 = \f{\rho^{\f79-\f2{d-1}}}{X^{1/3}}\ds^2_{(d+1\rightarrow 2)}\,,\qquad \e^{2A}= \f{\rho^{\f79-\f2{d-1}}}{X^{1/3}}\e^{2\cA}\,.
\end{equation}
The left-hand side in the expressions above denotes the quantities from the two-dimensional supergravity model while the right-hand side corresponds to the analytically continued expressions of the spherical brane construction in the $d \to 1$ limit. We thus find that the two metrics are  related by a simple conformal transformation. Carefully performing this conformal transformation and subsequent $d\to 1$ limit, one can show that the full action of the spherical brane model \eqref{eq:spherical-action} reduces to the action \eqref{eq:action-0} of the two-dimensional supergravity.

%%%%%%%%%%%%%%
\subsubsection{UV and IR analysis}
%%%%%%%%%%%%%%

For generic values of $d$, the UV and IR are located at 
\begin{equation}
	(X,Y)_{\rm UV} =(1, 0)\,,\qquad \text{and}\qquad  (X,Y)_{\rm IR} = \left(\f{d-1}3, \f{2(4-d)}3\right)\,.
\end{equation}
The UV locus is determined by the requirement that the spherical brane solution in this limit asymptotes to the usual near horizon geometry of coincident D$p$-branes with a flat worldvolume. The IR region is determined by the requirement that the metric smoothly caps off as $\dd r^2 +r^2 \dd\Omega^{2}_{d}$. This behaviour is the supergravity manifestation of the fact that at energies below the scale set by the radius of $S^d$ the dynamics of the MSYM theory is trivial, i.e. the radius of $S^d$ provides a natural IR cut-off.

A first step in analysing the behaviour of the solutions to the BPS equation is to expand them around the IR and UV loci. Before doing so let us note that for each $d<4$ there are two analytic solutions given by\footnote{Similar analytic solutions exist for $d>4$ but in this case the BPS equations are slightly modified and we will not present the solutions here. See \cite{Bobev:2018ugk} for details.}
\begin{equation}\label{eq:A1soln}
	\begin{aligned}
		Y^2 &= (3-d+X)^2\,,\qquad	\eta &= \eta_{0}\,,
	\end{aligned}
\end{equation}
and
\begin{equation}\label{eq:A2soln}
	\begin{aligned}
		Y^2 &= \f{X-1}{X}(3-d+X)^2\,,\\
		\e^{\f{2(d-1)(6-d)}{7-d}\left(\eta-\eta_{0}\right)} &= \f{(4-d)^{d-4}(d-1)^{6-d}}{36}(3-d+3X)^2 \,(1-X)^{4-d}\, X^{d-6}\,,
	\end{aligned}
\end{equation}
both containing a single integration constant $\eta_{0}$. Unfortunately, for general $d$ neither of these analytic solutions connect the IR and UV regions. For generic $d$ the first solution \eqref{eq:A1soln} reaches the desired UV but not the IR, while the second one \eqref{eq:A2soln} connects to the desired IR but does not reach the UV. Note that for $d=2$ the solution in \eqref{eq:A2soln} reaches both the UV and IR regions.

\paragraph{IR region} For generic $d$ we find two possible IR expansions. One of them can be resummed to the second analytic solution \eqref{eq:A2soln}, while the other one gives rise to the following behaviour for the scalars $Y^2$ and $\eta$,
\begin{equation}\label{eq:IR-exp}
	\begin{aligned}
		Y^2 &= \f49 \,(4-d)^2 + \f{2(d-4)(d+5)}{3(d-1)}\left(X-\f {d-1}3\right) + \cdots\,,\\
		\eta &= \eta_{\rm IR} - \f{3(7-d)(d+1)}{4(d-1)^2}\left(X-\f {d-1}3\right) + \cdots \,.    
	\end{aligned}
\end{equation}

\paragraph{UV region} In the UV region one also finds different branches of solutions to the linearised BPS equations. One of these branches can be resummed to yield the first analytic solution in \eqref{eq:A1soln}. The other branch is less trivial and leads to the following UV expansion for $Y^2$,
\begin{equation}
	\begin{aligned}
		Y^2 = Y_{\rm UV}^2 (X-1)^{\f{6-d}{2}}\Bigg[ & 1 + \f{8-(7-d)d}{2(d-4)}(X-1) + \f{d-6}{2(d-4)^2}\, Y_{\rm UV}^2\, (X-1)^{\f{4-d}{2}} \\
		& + \f18\left( (7-d)(1-d)-\f{8(d-5)}{(4-d)^2} \right)(X-1)^2 + \cdots \Bigg],.
	\end{aligned}
\end{equation}
This expansion is valid for $d<4$ but it should be treated with some care since for $d\geq 2$ the role of leading and sub-leading terms is exchanged. In order to study the UV expansion of the scalar $\eta$ it is convenient to first define,
\begin{equation}
	\label{eq:xi-def}
	\xi= \e^{- (d-1) a \eta} = \rho^{2a}\,,
\end{equation}
where $a=\f{4-d}{7-d}$, for which the BPS equation becomes,
\begin{equation}
	\f{\xi^\prime}{\xi} = \f{(d-4)((3-d+X)^2)-Y^2)}{2X(2(X-1)(3-d+X)+Y^2)}\,.
\end{equation}
The new scalar $\xi$ admits the following UV expansion,
\begin{equation}
	\begin{aligned}
		\xi = \xi_{\rm UV} (1-X)^{-\f14(4-d)^2}\Bigg[ & 1-\f14 (4-d)(3-d)(1-X) - \f{Y_{\rm UV}^2}{4}(2-d)^{\f{4-d}{2}}\\
		&+\f18\left((7-d)(1-d)-\f{8(d-5)}{(4-d)^2}\right)(1-X)^2 +\cdots \Bigg]\,.
	\end{aligned}
\end{equation}
We can continue the UV expansion indefinitely and find that it is completely fixed in terms of two integration constants, $\eta_{\rm UV}$, or equivalently $\xi_{\rm UV}$, and $Y_{\rm UV}$. For general $d<4$ the structure of the expansion is as follows,
\begin{equation}\label{eq:UV-exp-gen-p}
	Y^2 = Y_{\rm UV}^2\,(X-1)^{\f{6-d}{2}}\sum_{n=0}^\infty Y_{\rm UV}^{2n}\, (X-1)^{\f{(4-d)n}{2}} \sum_{k=0}^\infty a_{n,k}(X-1)^{k}\,,
\end{equation}
and similarly for $\xi$. We see that for $d\geq2$ the third term in the UV expansion becomes more leading than the second and so for these values of $d$ the series expansion should be rearranged. It is expected that the coefficients in this UV expansion should fix the sources and the vevs for the operators \eqref{eq:dual-ops} in the dual QFT. We will not attempt to make this holographic identification more precise here.

While the UV and IR expansions described above provide analytical control over the BPS equations, we were not able to find an analytic solution of these equations that connects these two expansions. Instead, we resort to numerics and find numerical solutions connecting the UV and the IR. In Figure~\ref{fig:BPSregions} we display several representative solutions for the function $Y(X)$ for some specific values of $d$ that illustrate the approach to $d \to 1$. In Figure~\ref{fig:BPSregiond16} we show a specific solution that illustrates the generic behaviour for $1<d<4$. The explicit numerical solutions fix the UV series expansion integration constants $\xi_{\rm UV}$ and $Y_{\rm UV}$ in terms of the IR integration constant $\eta_{\rm IR}$. This predicament is familiar from the spherical brane solutions for integer $d$ in \cite{Bobev:2018ugk} and other similar setups in Euclidean holography. We therefore find that for any real $1<d<4$ there is a numerical solution of the BPS equations which connects the UV and IR loci and depends solely on the integration constant $\eta_{\rm IR}$. In the IR region the solution is regular and the metric caps off smoothly as $\dd r^2 + r^2 \dd\Omega_d^2$, with $\dd\Omega_d^2$ the metric on the round~$S^d$. The circular D0-brane solution of interest is obtained by carefully taking the $d\to 1$ limit of this family of numerical solutions. While, strictly speaking, one finds a singular background for $d=1$ the limiting procedure described above provides a concrete method to extract finite holographic observables from this supergravity background as we show in detail in Section~\ref{sec:FandWL} below. We now proceed with a discussion on how to uplift this limiting $d=1$ solution to 10d and 11d supergravity.

%%%%%%%%%%%
\subsection{Uplift to type IIA and eleven-dimensional supergravity}
\label{subsec:10dand11d}
%%%%%%%%%%%

The 2d gravitational model in \eqref{eq:action-0} is obtained as a consistent truncation of the maximal $\so(9)$ gauged supergravity theory which in turn can be obtained from a consistent truncation of the 10d type IIA supergravity on $S^8$ or 11d supergravity on $S^1\times S^8$, see \cite{Bossard:2022wvi} and \cite{Bossard:2023jid} for a recent discussion. After analytically continuing the $S^8$ to $\dS_8$, this sequence of consistent truncations implies that the circular D0-brane solution constructed above can be uplifted to a solution of 10d or 11d supergravity. Here we perform this uplift explicitly.

Rather than resorting to the uplift formulae derived in \cite{Bossard:2022wvi,Bossard:2023jid} using exceptional field theory, we construct the 10d supergravity background by relying on the familiar structure of the spherical D$p$-brane solutions in \cite{Bobev:2018ugk} to which we refer the reader for more details as well as a precise definition of our conventions. The metric (in string frame) and the 10d dilaton can be expressed as 
\begin{equation}
	\begin{aligned}
		\ds_{10}^2 =&\, \rho^{-1/3} Q^{-1/2}\left[ X^{1/3}\rho^{-4/9}\ds_2^2 + \f{1}{g^2} \left( \dd\theta^2 + P \cos^2\theta \dd\tilde\Omega_2^2 + Q \sin^2\theta \dd\Omega_5^2 \right)\right]\,,\\
		\e^\Phi =&\, g_s\rho^{-7/6}P^{1/2}Q^{1/4}\,,
	\end{aligned}
\end{equation}
where the functions $P$ and $Q$ are defined as,
\begin{equation}
	P = \f{X}{X\sin^2 \theta + (X^2-Y^2)\cos^2\theta}\,,\qquad Q = \f{X}{\sin^2\theta + X\cos^2\theta}\,. 
\end{equation}
In these expressions, $\dd\Omega^2_n$ denotes the metric on the (unit radius) round $n$-sphere with volume form $\dvol_n$, $\dd\tilde\Omega_2^2$ is the metric on unit radius $\dS_2$, and $\theta\in\left[0,\f\pi 2\right]$. In terms of these functions, the NSNS- and RR-potentials can be written as
\begin{equation}
	\begin{aligned}
		B_2 =&\, \f{YP}{g^2 X\rho^{1/3}}\cos^3\theta\,\wti\dvol_2\,,\\
		C_5 =&\, \f{\ii YQ \rho^{1/3} }{g_s g^5 X}\sin^4\theta \,\dvol_5\,,\\
		C_7 =&\, \f{\ii}{g_s g^7} \left( \omega(\theta) + P \cos\theta\sin^6\theta\right)\wti\dvol_2\wedge\dvol_5\,,
	\end{aligned}
\end{equation}
where $\omega(\theta)$ is defined such that $\dd(\omega(\theta) + \cos\theta\sin^6\theta)\wedge \wti\dvol_2\wedge\dvol_5= 7 \wti\dvol_8$ is proportional to the volume form on $\dS_8$. Note that with the identification \eqref{eq:maptorho}, this reduces exactly to the uplift for the spherical brane solutions \eqref{eq:spherical-metric}-\eqref{eq:spherical-forms}. It will prove convenient to also have explicit expressions for the dual RR-potentials which take the form
\begin{equation}
	\begin{aligned}
		C_1 =&\,-\f{\ii \rho X}{g_sgY} \left( P^{-1} + 2 Q^{-1} \right)\dd\tau\,,\\
		C_3 =&\,-\f{2\ii\rho^{2/3}P}{g_sg^3 Q}\cos^3\theta\dd\tau\wedge\wti\dvol_2\,.
	\end{aligned}
\end{equation}
The associated field strengths are defined as 
\begin{equation}
	H_3 = \dd B_2\,, \qquad F_p = \dd C_{p-1} - H_3\wedge C_{p-3}\,,
\end{equation}
and satisfy the following duality relations,
\begin{equation}
	F_2 = \star F_8\,,\qquad F_4 = \star F_6\,.
\end{equation}
We have checked explicitly that the two-dimensional BPS equations imply that the 10d background above solves the equations of motion of ten-dimensional type IIA supergravity. 

The conserved (Page) D0-brane charge can be obtained by evaluating the following integral
\begin{equation}
	N = \f{1}{(2\pi\ell_s)^7} \int_{\cM_8} \dd C_7 = \f{7\vol_8}{(2\pi \ell_s g)^7}\,,
\end{equation}
where $\vol_8$ is the regularised volume of the unit radius $\dS_8$ which is nothing but the volume of $S^8$, see \eqref{eq:defVd}.\footnote{In practice we compute the flux by analytically continuing the setup to purely Euclidean signature and performing the integrals there. This procedure naturally gives rise to sphere volumes.} Since the Page charge is conserved, we can choose to evaluate this integral in the UV region. 

The 10d background above can be further uplifted to 11d supergravity, where the metric and three-form potential take the form,
\begin{equation}
	\begin{aligned}
		\ds_{11}^2 =&\, P^{-1/3} Q^{-2/3}\Bigg[ X^{1/3} \ds_2^2 + \f{\rho^{4/9}}{g^2}\left(\dd\theta^2 + P\cos^2\theta \dd\wti\Omega_2^2 + Q \sin^2\theta \dd\Omega_5^2 \right) \\
		&\,+ \f{\rho^{-14/9}PQ}{gg_s}\left( gg_s \,\dd x_{11} + \f{\rho X}{Y}(P^{-1}+2Q^{-1})\dd\tau \right)^2\Bigg]\,,\\
		A_3 =&\, - \f{PY \cos^3\theta}{g^3g_s X\rho^{1/3}}\left( gg_s \,\dd x_{11} + \f{2\rho X}{YQ}\dd\tau \right)\wedge \wti\dvol_2\,.
	\end{aligned}
\end{equation}
%

%%%%%%%%%%%
\subsection{Embedding in LM}
%%%%%%%%%%%

In \cite{Lin:2005nh} Lin and Maldacena constructed the gravitational type IIA duals to field theories with 16 supercharges and $\su(2|4)$ global symmetry, such as the vacua of the Lorentzian BMN matrix quantum mechanics by dimensionally reducing the M-theory solutions in \cite{Lin:2004nb}. Before discussing the analytic continuation of their setup to Euclidean signature, let us briefly review the Lorentzian case.

The bosonic part of the $\su(2|4)$ symmetry algebra is $\su(2)\times \su(4)$, which is realised geometrically by the Killing isometries associated with a two- and a five-sphere in the metric. Additionally to these, the background of \cite{Lin:2005nh} exhibits time translation symmetry. Due to the large amount of supersymmetry the full supergravity solution can be obtained from a single function $V$ depending on two coordinates $s$ and $z$. The 10d supergravity BPS equations furthermore imply that this function is given by an axisymmetric solution to the 3d Laplace equation,
\begin{equation}\label{eq:LM-Laplace}
	\ddot V + s^2 V^{\prime\prime} = 0\,,
\end{equation}
where dots indicate derivatives with respect to $\log s$ and primes derivatives with respect to $z$. Any solution to this equation has to be supplemented with appropriate boundary conditions that ensure the regularity of the supergravity background.

\begin{figure}[!htb]
	\centering
	\begin{tikzpicture}
		\node at (0,0) {\includegraphics[width=0.7\textwidth]{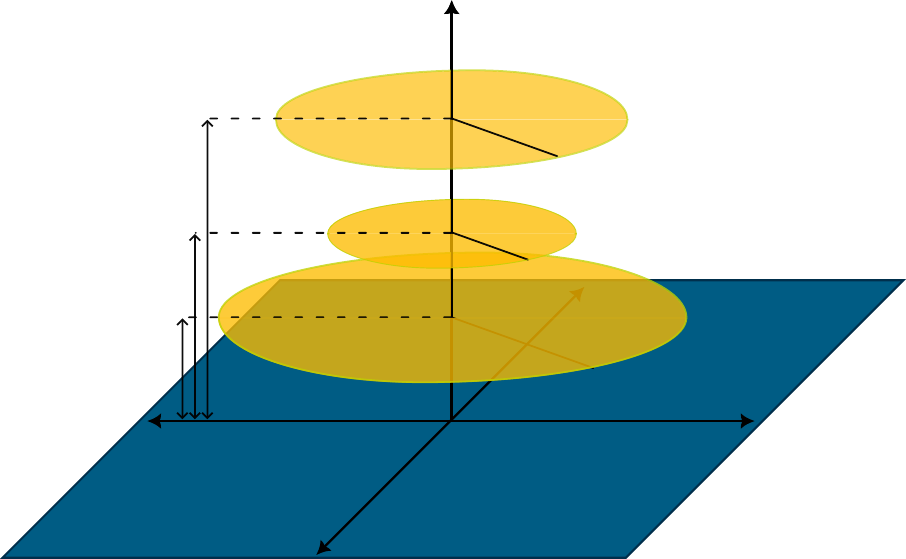}};
		\node at (3.8,-1.5) {$s$};
		\node at (0.3,3.3) {$z$};
		\node at (-3.6,-1) { $z_1$};
		\node at (-3.45,0.05) {$z_2$};
		\node at (-3.3,1.2) {$z_3$};
		\node at (1.6,2.05) {$\mathbb{D}_1$};
		\node at (1.1,0.55) {$\mathbb{D}_2$};
		\node at (2.3,-0.45) {$\mathbb{D}_3$};
		\node at (0.8,-1) {$r_1$};
		\node at (0.2,0.3) {$r_2$};
		\node at (0.5,1.6) {$r_3$};
	\end{tikzpicture}
	\caption{An illustrative configuration for the electrostatic problem defining a solution of the form \eqref{eq:LM-sol}. The full setup is rotationally symmetric and the angular coordinate in the $s$-plane is not part of the $10d$ solution.}
	\label{fig:LM-discs}
\end{figure}
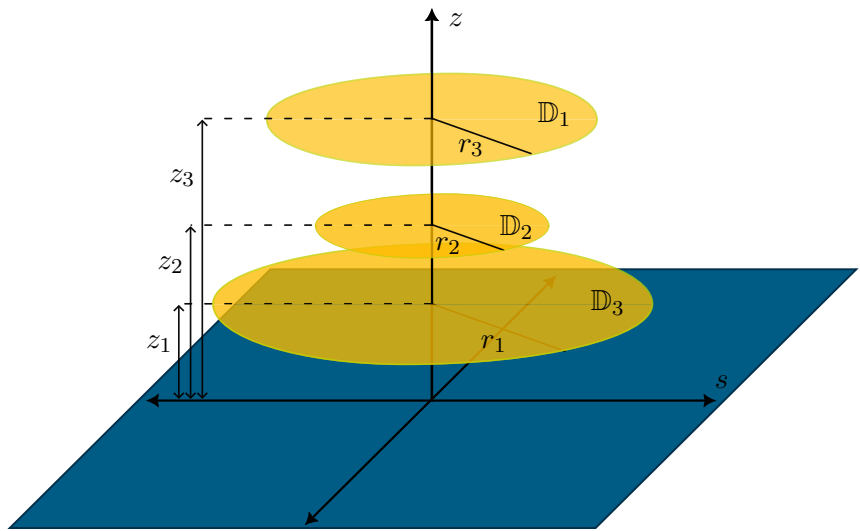

The type IIA supergravity solution can then be written as,\footnote{We have introduced $g_s$ and changed signs of some of the form fields to match with our conventions.}
\begin{equation}
	\label{eq:LM-sol}
	\begin{aligned}
		\ds_{10}^2 =&\, \left( \f{\ddot{V}-2\dot V}{-V^{\prime\prime}}\right)^{1/2}\left[ - \f{4\ddot V}{\ddot V-2\dot V}\dd t^2 + \f{-2V^{\prime\prime}}{\dot V}(\dd s^2 + \dd z^2) + 4 \dd\Omega_5^2 + 2\f{V^{\prime\prime}\dot V}{\Delta}\dd\Omega_2^2 \right]\,,\\
		\e^{4\Phi} =&\, \f{4g_s^4(2\dot V-\ddot V)^3}{V^{\prime\prime}\dot V^2 \Delta^2}\,,\qquad B_2 =\, -2\left( \f{\dot V\dot V^{\prime}}{\Delta}+z \right)\dvol_2\,,\\
		C_1 =&\, -\f{2\dot V\dot V^\prime}{g_s(2\dot V-\ddot V)}\dd t\,, \qquad	C_3 =\, -\f{4\dot V^2 V^{\prime\prime}}{g_s\Delta} \dd t\wedge \dvol_2\,,
	\end{aligned}
\end{equation}
where we defined the function $\Delta = (\ddot V-2\dot V)V^{\prime\prime} - \dot V^{\prime 2}$ and $\dvol_2$ is the volume form on $\dd\Omega_2^2$. The metric represents an $S^2\times S^5$ fibration over the $(s,z)$ plane. The five-sphere shrinks to zero size on the $s=0$ axis, while the $S^2$ shrinks to zero size when $\f1s \dot V=0$. Different supergravity solutions are therefore specified by different boundary conditions for the Laplace equation. As explained in \cite{Lin:2005nh}, the appropriate boundary conditions to describe the vacua of the BMN matrix quantum mechanics are given by inserting a collection of discs $\disc_i$ with radii $r_i$ at specific values $z_i$ parallel to the $s$-axis, see Figure \ref{fig:LM-discs}.

Corresponding to each choice of parameters we have an electrostatics problem whose solution results in a supergravity equation of the form \eqref{eq:LM-sol}. In addition to imposing that the $S^2$ smoothly shrinks at the location of the discs, the parameters are constrained by demanding that the fluxes through each non-contractible cycle are appropriately quantised. In particular, all vacua of the BMN matrix quantum mechanics have an electrostatic system with an infinite conducting plate at $z=0$ where the potential vanishes. Furthermore, in order for the solution to asymptotically approach the supergravity solution produced by flat coincident D0-branes we need to turn on the following background potential
\begin{equation}
	\label{eq:V-background}
	V_{\infty} = V_0\left( s^2 z- \f23 z^3 \right)\,. 
\end{equation}
The supersymmetric vacua of the BMN matrix quantum mechanics are then mapped to collections of discs as follows. Each vacuum is defined by a particular (not necessarily reducible) representation $\mathfrak{R}$ of $\su(2)$. If the $d_i$-dimensional representation appears $n_i$ times, we have to put a conducting disc with charge $n_i$ at $z_i = d_i$. In order for the supergravity solution to be non-singular, the radius of the disc is determined in terms of the charge, such that the charge density vanishes at the edge of the disc. This precisely mimics the QFT discussion on the vacuum moduli space in Section \ref{sec:BMN}.

The fluxes should be appropriately quantised along all non-contractible cycles in this geometry and are in one-to-one correspondence with the quantum mechanics vacua. In particular, since the $S^5$ shrinks on the $s=0$ axis, we can form topologically $S^6$ cycles, starting on the axis, going around a number of discs and coming back to the axis, carrying D2-brane flux $Q_j$. Similarly, the $S^2$ shrinks at the location of the discs so we can form topologically $S^3$ cycles, starting at a disc and ending at another disc, carrying NS5-brane flux $S_k$. From this setup we can form a Young tableau by adding $Q_j$ rows of length $j$ where $\sum Q_j = N$. The NS5-brane fluxes correspond in a similar way to the dual Young tableau. 

As discussed in Section \ref{sec:BMN} there are two ways to preserve supersymmetry in the Euclidean BMN model. In the supergravity description, the first one corresponds to analytically continuing the $S^5$ to $\dS_5$. In this case the analysis is entirely analogous to the Lorentzian model described above and we will not discuss it further. Alternatively, one can analytically continue the $S^2$ to $\dS_2$. This is precisely the setup we arrive at when approaching the BMN matrix quantum mechanics from MSYM on $S^d$ as we now describe. The full solution is still completely determined as a function of $V$ which solves the equation \eqref{eq:LM-Laplace}, and can be obtained by simply replacing 
\begin{equation}
\dd\Omega_2^2 \rightarrow \dd\wti\Omega_2^2\,,\qquad \dvol_2 \rightarrow \ii\, \wti\dvol_2\,.
\end{equation}
We do not have to explicitly analytically continue the time-direction $t$ but instead the metric function $g_{tt}$ itself changes sign as we will explain. Given the properties of our new circular D0-brane solution we conjecture that it is dual to the unique vacuum of the relevant Euclidean BMN matrix quantum mechanics which corresponds to a single disc with charge $1$. To this end it is instructive to understand how the circular D0-brane solution in Section~\ref{subsec:10dand11d} can be embedded in the general ansatz \eqref{eq:LM-sol}. We find that indeed the solution in Section~\ref{subsec:10dand11d} can be written as an LM background where the function $V$ is given by
\begin{equation}
	V = -\rho^{2/3}\f{\cos\theta\left( X-2 + \left(3X+2\right) \cos 2\theta \right) }{48 g^3 X} \,,
\end{equation}
while the LM coordinates $z$ and $s^2$ are defined through the coordinate change
\begin{equation}\label{eq:sandz}
	z =\, -\ii\f{(2+X)}{4g^2Y\rho^{1/3}}\cos\theta\,, \qquad\qquad s^2 =\, -\f{X((2+X)^2-Y^2)}{16g^4 Y^2 \rho^{2/3}}\sin^2\theta\,.
\end{equation}
Note that even though the coordinates $z$ and $s$ are seemingly both analytically continued to timelike directions, the various metric factors in the LM metric determined in terms of the function $V$ and its derivative ensure that the metric is of signature $(1,9)$ as expected. 
In order to compare with the LM metric in \eqref{eq:LM-sol}, we explicitly list the following derivatives of $V$ as a function of $X$,
\begin{equation}\label{eq:LM-map}
	\begin{aligned}
		\dot V =&\, \f{\rho^{2/3}}{4g^3}\cos\theta \sin^2\theta\,,\qquad &
		\ddot V =&\, \f{((2+X)^2 -Y^2)P \rho^{2/3}}{2g^3(4(1+X)P+X Q)}\sin^2\theta\cos\theta\,,\\
		\dot V^\prime =&\, \f{\ii}{g} \f{(2P+Q)Y\rho}{4(1+X)P+X Q}\sin^2\theta\,,\qquad &
		V^{\prime\prime} =&\, \f{8gPY^2\rho^{4/3}}{X(4(1+X)P+XQ)}\cos\theta\,.
	\end{aligned}
\end{equation}
In the UV the circular D0-brane solution approaches the flat D0-brane background. In the LM coordinates, the UV is located at large $z$ and $s$, where this asymptotic behaviour is reflected in the background potential \eqref{eq:V-background}. On the other hand, in the IR, the circular D0-brane solution is singular. This is perhaps due to the fact that it arises as an uplift of a gauged supergravity solution and it is notoriously hard to describe localised brane sources using gauged supergravity. Analysing where the $S^2$ shrinks, or equivalently where $P\cos^2\theta$ vanishes, we find an infinitely charged disc at $z=0$. The finite charged disc at a small distance from the $z=0$ plane is absent in our description. However, going to the IR we find that $P\rightarrow 0$ causing the two-sphere to shrink on the interval\footnote{As will be discussed in the next section, the integration constant $\rho_{\rm IR}$ blows up in the strict $d\to 1$ limit so that this smeared density collapses to the $z=0$ plane.}
\begin{equation}
	z\in \left[0,\f{1}{4g^3\rho_{IR}^{1/3}}\right]\,,
\end{equation}
where we find a continuous charge density. Finally, in line with the expectation from the LM setup, we find that the size of the $S^5$ appropriately shrinks to zero size at $s=0$. This analysis shows that the circular D0-brane solution shares various crucial properties with the trivial vacuum state of the BMN matrix quantum mechanics and therefore lends further evidence to our holographic interpretation.

%%%%%%%%%%%%%%%%%%%%%%%%%%%%%%%%%
\section{Holographic free energy and Wilson loop vev}          
\label{sec:FandWL}	    
%%%%%%%%%%%%%%%%%%%%%%%%%%%%%%%%%

It is natural to conjecture that the supergravity solutions found above are holographically dual to the MSYM theory on $S^d$ as discussed in Section~\ref{sec:BMN}. Since both the supergravity and the field theory constructions admit an analytic continuation in the number of dimensions $d$ we can expect that this holographic correspondence can be established for any value of $d$. In particular, we should be able to take the $d \to 1$ limit and elucidate aspects of the physics of the BMN matrix quantum mechanics. The goal of this section is to show that the supersymmetric localisation results for the $S^d$ free energy and the supersymmetric Wilson loop vev in Section~\ref{sec:BMN} agree with the appropriate dual supergravity observables, namely the regularised supergravity on-shell action and the on-shell action of a probe string suitably embedded in the spherical brane background.

%%%%%%%%%%%
\subsection{Free energy}
\label{sec:holofreeenergy}
%%%%%%%%%%%

The holographic free energy of the MSYM theory on $S^d$ can be obtained by computing the appropriately regularised on-shell action of the $(d+1)$-dimensional supergravity solution. We calculate this on-shell action for any $d$ and pay particular attention to the subtleties arising in the $d \to 1$ limit. Since the solutions are spherically symmetric the calculation of the on-shell action is in principle straightforward and involves a single integral along the radial coordinate. Nonetheless, this is technically challenging since we are not able to  solve the BPS equations analytically for general $d$ and have to resort to a numerical analysis. Fortunately, as we show below, this challenge can be largely circumvented by noting that the supergravity on-shell action can be written as a total derivative. 

To this end we employ the so-called scaling similarity of the supergravity theory. If one rescales the dilaton scalar and the metric with any real number $t$ in the following way
\begin{equation}
	\label{eq:scalingtransformation}
	\e^{\f{d-1}{d-7}\eta} \rightarrow t \e^{\f{d-1}{d-7}\eta} \,,\qquad\qquad g_{\mu\nu}\to t^{-2+\f6{d-1}}g_{\mu\nu}\,,
\end{equation}
while keeping the scalars $X$ and $Y$ invariant one finds that the supergravity equations of motion and BPS equations are invariant but the action \eqref{eq:spherical-action} is rescaled by a constant,
\begin{equation}
	\label{eq:scaling}
	S \to t^{4-d} S\,.
\end{equation}
Taking a derivative with respect to the parameter $t$ and subsequently setting it to zero then implies,
\begin{equation}
	\label{eq:scaling-trick}
	(4-d)S = \f{\dd}{\dd t}t^{4-d}S\Big|_{t\to 1} = \f{\dd S_t}{\dd t}\Big|_{t\to 1}\,,
\end{equation}
where $S_t$ denotes the action with all fields replaced by their rescaled values \eqref{eq:scalingtransformation}. Writing the rescaled fields as $\phi_t$, for a given field $\phi$, we can use the chain rule to compute the derivative in~\eqref{eq:scaling-trick}. Defining the canonical momenta $P_\phi^\mu = \f{\delta \cL}{\delta (\partial_\mu \phi)}$, we find
\begin{equation}
	%\begin{aligned}
			(4-d)S =\, \sum_\phi \int \left[ \f{\delta \cL}{\delta\phi}\f{\dd \phi_t}{\dd t} + P^\mu_\phi \partial_\mu \f{\dd \phi_t}{\dd t} \right]_{t\to 1}
			=\, \sum_\phi \int \partial_\mu\left[ P^\mu_\phi  \f{\dd \phi_t}{\dd t} \right]_{t\to 1}\,,
	%\end{aligned}
\end{equation}
where to arrive at the second identity we used the equations of motion. We can further simplify this expression by using the spherical symmetry of the solutions of interest and write the action in terms of a one-dimensional Lagrangian. To do this it is useful to introduce an auxiliary field $B$ in the metric,
\begin{equation}
	\ds_{d+1}^2 = \e^{2B}\dd r^2 + \cR^2 \e^{2\cA}\dd\Omega_d^2\,,
\end{equation}
such that both $A$ and $B$ transform in the same way under the scaling transformation \eqref{eq:scalingtransformation}. The momentum $P_B$ is the one-dimensional Hamiltonian which vanishes on-shell, so we can immediately eliminate it. We then find,
\begin{equation}\label{eq:Stotder}
	(4-d)S = \int \f{\dd}{\dd r}\left[ \f{4-d}{d-1}P_\cA + \f{d-7}{d-1}P_\eta \right]\dd r\,.
\end{equation}
That is the one-dimensional Lagrangian is a total derivative and the action can be reduced to surface terms.
Using the BPS equations and rewriting the expression above in terms of the radial coordinate $X$ one finds
\begin{equation}
	\label{eq:total-der-S}
	\begin{aligned}
		S  =\, \int \f{\dvol_{d}\wedge \dd X}{2\kappa_{d+1}^2}\f{\dd}{\dd X}\bigg[ \f{\cR^{d+1}\e^{(d+1)\cA}}{Y}\bigg( Y^2 \left( \f{d-1}{d-4}\f{1}{X} +3 \right)- (3-d+X)(6-d+3X) \bigg) \Bigg]\,.	
	\end{aligned}	 
\end{equation}
Since the on-shell action is an integral of a total derivative one can evaluate it using the UV and IR expansion without the need to construct a full numerical solution of the non-linear BPS equations and integrate over the radial direction. Nevertheless, the existence and explicit construction of this numerical solution are needed in the holographic context since this is the mechanism to establish a relation between the UV and IR integration constants which in turn enter in the holographic dictionary. We will describe how this works in our setup in more detail below.

As usual, the supergravity action is divergent in the UV and needs to be appropriately regularised. Due to the non-conformal nature of our setup, the holographic renormalisation procedure is more subtle but, as explained in \cite{Kanitscheider:2008kd,Bobev:2019bvq}, can be implemented most clearly in the so-called ``dual frame''. To define the dual frame one needs to perform the following conformal transformation,
\begin{equation}
	g_{\mu\nu} \rightarrow \wti g_{\mu\nu} = \e^{2a \eta}g_{\mu\nu}\,,\qquad \qquad 	a = \f{4-d}{7-d}\,,
\end{equation}
where the constant $a$ was already defined in \eqref{eq:xi-def}. This frame is characterised by the property that it is invariant under the scaling transformation \eqref{eq:scaling} or equivalently, that the dual frame metric does not contain explicit dependence on the dilaton $\eta$ (or $\rho$ through \eqref{eq:maptorho}).
The convenient feature of the dual frame is that the counterterms needed for holographic renormalisation take a very similar form to those used for asymptotically locally AdS spaces. 

In addition to the Gibbons-Hawking boundary term, there are infinite counterterms that need to be added to render the on-shell action finite. We have to add a ``superpotential counterterm''
\begin{equation}\label{eq:Ssuperpot}
	S_{\rm superpot} = -\f{1}{2\kappa_{d+1}^2} \int \dd^{d}x \sqrt{\wti h}\e^{-a\, d\eta} \sqrt{\e^\cK \cW\wti\cW}\Big|_{Y\rightarrow 0}\,. 
\end{equation}
where $\cW$ and $\wti\cW$ are defined in Appendix \ref{app:Spherical-p}. Note that we only add the part with $Y\rightarrow 0$, i.e. the part which is relevant for regularising the on-shell action of the flat D$p$-brane solutions appears. In the UV, as $X\rightarrow 1$ and $Y\rightarrow 0$ this counterterm reduces exactly to the cosmological constant counterterm. Since the asymptotic boundary of our solutions is not flat we also need to add the curvature counterterm \cite{Emparan:1999pm},
\begin{equation}\label{eq:Scurv}
	S_{\rm curv} = \f{1}{4g\kappa^2_{d+1}} \int\dd^{d}x \sqrt{\wti h} \e^{-a (d-1) \eta} R \big[\wti h\big] + \cdots\,,
\end{equation}
where the dots denote additional terms that are only relevant for $d > 4$. Since in this work we focus on $d<4$ we will ignore these henceforth. In addition to the terms above, there is an infinite counterterm originating from the presence of the scalar $Y$. This counterterm takes the form
\begin{equation}\label{eq:SY}
	S_{Y} = -\f1{4\kappa_{d+1}^2} \int \dd^{d}x\sqrt{\wti h}\e^{- a (d-1) \eta} Y^2\,.
\end{equation}
For the cases of interested here, i.e. $1<d<4$ the three terms in \eqref{eq:Ssuperpot}, \eqref{eq:Scurv}, and \eqref{eq:SY} are all the infinite counterterms. 

In addition to the infinite counterterms described above one should study all possible covariant finite counterterms that can be added. It turns out that for all $d<4$ there is a universal finite counterterm parametrising the coupling of the scalar $\log X$ to the curvature of the $d$-sphere,
\begin{equation}\label{eq:Sfinite}
	S_{\rm finite} = \f{1}{g\kappa_{d+1}^2}\int \dd^{d}x \sqrt{\tilde h}\e^{-a(d-1) \eta} R\big[\tilde h\big] \log X\,.
\end{equation}
Note that this universal counterterm was not presented in \cite{Bobev:2019bvq}. The universal nature of this finite counterterm suggests that it should be added to the evaluation of the on-shell action. Presumably, imposing that the holographic renormalisation procedure is compatible with supersymmetry will uniquely fix the coefficient of this counterterm. Since in this work we do not perform a proper supersymmetry analysis of the boundary terms in our supergravity model, we will keep the coefficient of $S_{\rm finite}$ arbitrary and use other arguments to fix it later.

Using the expression of the on-shell action as a total derivative, \eqref{eq:total-der-S}, and noting that the IR contribution completely vanishes, we find that the on-shell action is given by 
\begin{equation}\label{eq:OSAfinite}
	S\Big|_{\rm UV} + S_{\rm inf} + c(d) S_{\rm finite} = \left[f_1(d)+ c(d)f_2(d) \right]\f{\vol_{d}Y_{\rm UV}^{2-d}\,\xi_{\rm UV}}{\kappa_{d+1}^2g^{d-1}}\,,
\end{equation}
where $S_{\rm inf}$ is the sum of the three infinite counterterms in \eqref{eq:Ssuperpot}, \eqref{eq:Scurv}, and \eqref{eq:SY}. The constants $f_1(d)$ and $f_2(d)$ are the contributions obtained by evaluating respectively the action plus infinite counterterms and the finite counterterm in the UV, using the UV expansion derived in Section~\ref{subsec:circularD0}, and read
\begin{equation}\label{eq:f1-f2}
	\begin{aligned}
		f_1(d) =&\, \f18(4-d)^{d-3}\left( 5(4-d)^2 - 2(d-1)(6-d)(3-d) \right)\,,\\
		f_2(d) =&\, (4-d)^{d-2}d(d-1)\,.
	\end{aligned}
\end{equation}
The constant $c(d)$ is the coefficient of the finite counterterm in \eqref{eq:Sfinite} and will be fixed momentarily. Note that the solutions are labelled by a single integration constant $\eta_{\rm IR}$ where the UV constants $Y_{\rm UV}$ and $\xi_{\rm UV}$ should be expressed in terms of this single constant. 

\begin{figure}[!htb]
	\centering
	\begin{tikzpicture}
		\node at (0,0) {\includegraphics[width=0.7\textwidth]{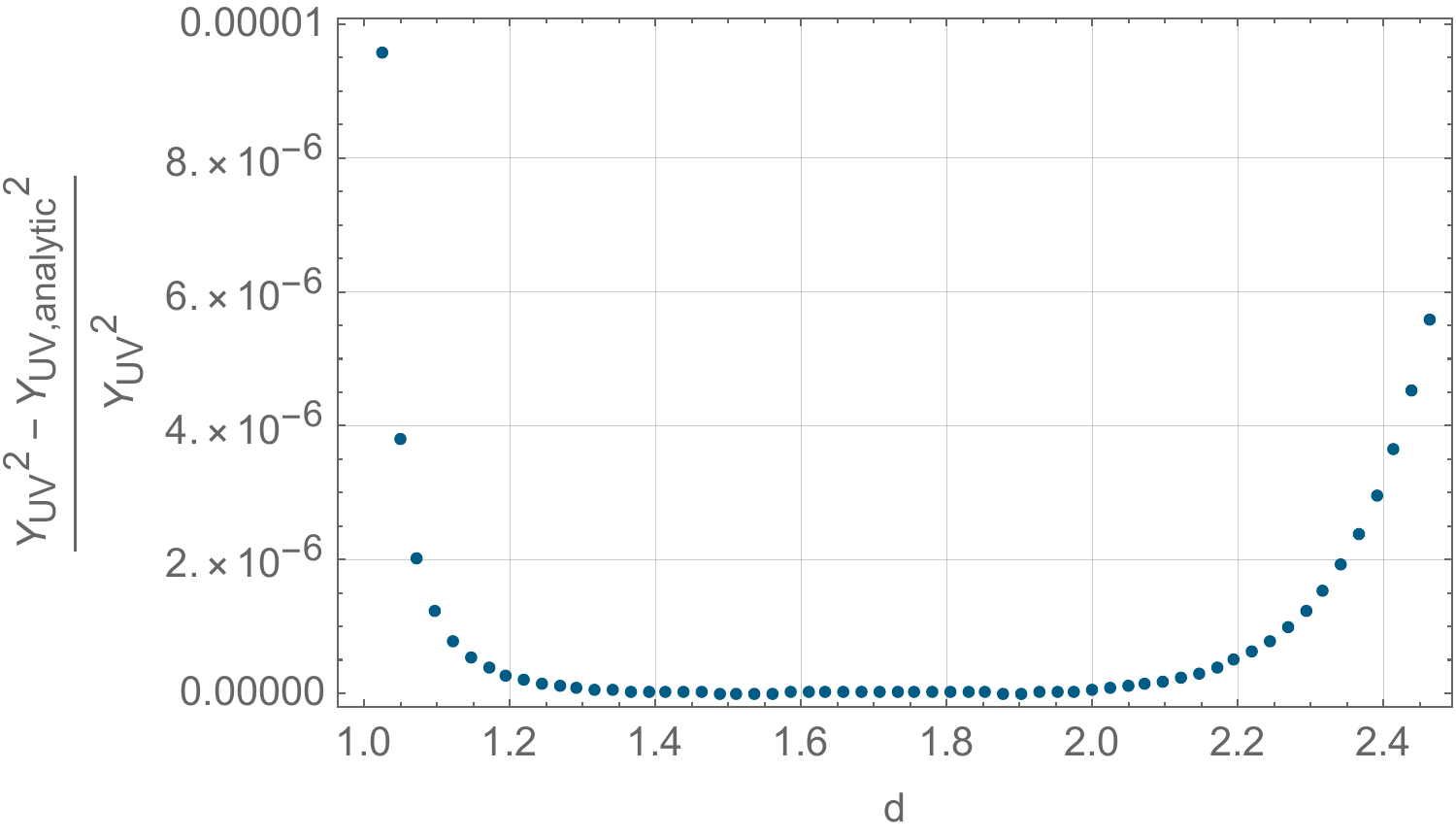}};
	\end{tikzpicture}
	\caption{The relative error for $Y_{\rm UV}^2$ compared to the analytic expression in \eqref{eq:YUVanalytic} as a function of the dimension $d$. For $d\lesssim 2.5$ the agreement between the analytic expression and the numerical results is excellent.}
	\label{fig:YUVfit}
\end{figure}

In order to compare the supergravity on-shell action with the QFT free energy presented in Section~\ref{sec:BMN} we need the following dictionary between the string theory and QFT parameters, see \cite{Bobev:2019bvq} and Appendix~\ref{app:Spherical-p},
\begin{equation}\label{eq:kappadef}
	\kappa_{d+1}^2 = \f{(2\pi\ell_s)^8g_s^2 \Gamma\left( \f{10-d}{2}\right) g^{9-d}}{8\pi^{\f{12-d}{2}}}\,,\qquad (2\pi \ell_s g)^{d-8} = \f{g_s N}{2\pi \vol_{7-d}}\,.
\end{equation}
In addition, we follow \cite{Bobev:2019bvq} and define the effective dimensionless 't~Hooft coupling as 
\begin{equation}
	\label{eq:lambda-hol}
	\lambda = \f{2\pi g_s N}{(2\pi\ell_s)^{4-d}}\cR^{4-d}\e^{(4-d)\cA}\e^{\f{10-d}{7-d}\eta}\Big|_{\rm UV} = \f{8\pi^{\f{12-d}{2}}(4-d)^{4-d}8}{(2\pi\ell_s g)^{2(6-d))}\Gamma\left( \f{8-d}{2} \right)} \, Y_{\rm UV}^{d-4}\xi_{\rm UV}^{-\f{6-d}{4-d}}\, \,.
\end{equation}
Plugging all this in the expression for the on-shell action in \eqref{eq:OSAfinite} we find the holographic result for the free energy of the MSYM theory on $S^d$. The resulting expression for the on-shell action is however not very useful, since it still contains the UV integration constants $\xi_{\rm UV}$ and $Y_{\rm UV}$ as well as the undetermined coefficient of the finite counterterm $c(d)$. To rectify this we can express $\xi_{\rm UV}$ in terms of $\lambda$ using \eqref{eq:lambda-hol}. Next, we are left to determine $Y_{\rm UV}$ as a function of $d$. To do so we numerically solve the BPS equations imposing the smooth IR boundary conditions and extract the UV integration constant by comparing the numerical solution with the UV expansion of the BPS equations. Based on the very accurate numerical results we find the following analytic expression for $Y_{\rm UV}$, 
\begin{equation}\label{eq:YUVanalytic}
	Y_{\rm UV}^2 = \f{(4-d)^2\sqrt{\pi}}{2\Gamma\left(\f{d-1}{2}\right)\Gamma\left(\f{6-d}{2}\right)}\,.
\end{equation}
This result agrees with the numerics to great accuracy with, for $d\lesssim 2.5$, a relative error $\delta Y_{\rm UV}^2 = (Y_{\rm UV, num.}^2-Y_{\rm UV, an.}^2)/Y_{\rm UV, num.}^2$ of around $10^{-6}$, see Figure \ref{fig:YUVfit}. For $d\gtrsim 2.5$ the fit of the numerical data to the analytic expression is somewhat less accurate. This can be attributed to the fact that as $d$ is increased the sub-leading terms in the UV expansion of $Y$ \eqref{eq:UV-exp-gen-p} become more important and have to be taken into account.

%%%%%%%%%%
%%%%%%%%%%

The last order of business is to find the value for the coefficient of the finite counterterm $c(d)$. We fix this coefficient as follows; we first fix the dimension to $d=2$ and $d=3$ and insist that the holographic on-shell action agrees with the supersymmetric localisation result for the QFT free energy as in \cite{Bobev:2019bvq}. We then impose that  $c(d)$ is a simple rational function of the dimension $d$. These two constraints are compatible with the following expression for $c(d)$
\begin{equation}\label{eq:c-anal}
	c(d) = \f{5}{8}\f{4-d}{d(d-1)}\,.
\end{equation}
With this choice of the coefficient of the finite counterterm and using the expressions for the on-shell action in \eqref{eq:OSAfinite}, together with \eqref{eq:f1-f2}, \eqref{eq:kappadef}, \eqref{eq:lambda-hol}, and \eqref{eq:YUVanalytic}, we find the following result for the holographic free energy of the spherical D-brane solutions
\begin{equation}
	\label{eq:F-exact-d}
	\f{F_d}{N^2} = - \f{16\pi^{\f{(d+1)(4-d)}{2(6-d)}}(6-d)}{\lambda\,\Gamma\left( \f{d-3}{2} \right)(8-d)(d-4)}\left[ \f{\lambda}{4}\Gamma\left( \f{8-d}{2} \right)\Gamma\left( \f{6-d}{2} \right)\Gamma\left( \f{d-1}{2} \right) \right]^{\f{2}{6-d}}\,.
\end{equation}
This expression matches exactly the analytically continued expression of the free energy obtained from the large $N$ and large $\lambda$ limit in the supersymmetric localisation matrix model as described in \cite{Bobev:2019bvq} and Section~\ref{sec:BMN} above. This result amounts to a precision test of our proposed holographic duality and in particular facilitates taking the $d \to 1$ limit appropriate for circular D0-branes.

%%%%%%%%%%%
\subsection{Wilson loop vev}
\label{subsec:WLvev}
%%%%%%%%%%%

\begin{figure}[!htb]
	\centering
	\begin{tikzpicture}
		\node at (0,0) {\includegraphics[width=0.7\textwidth]{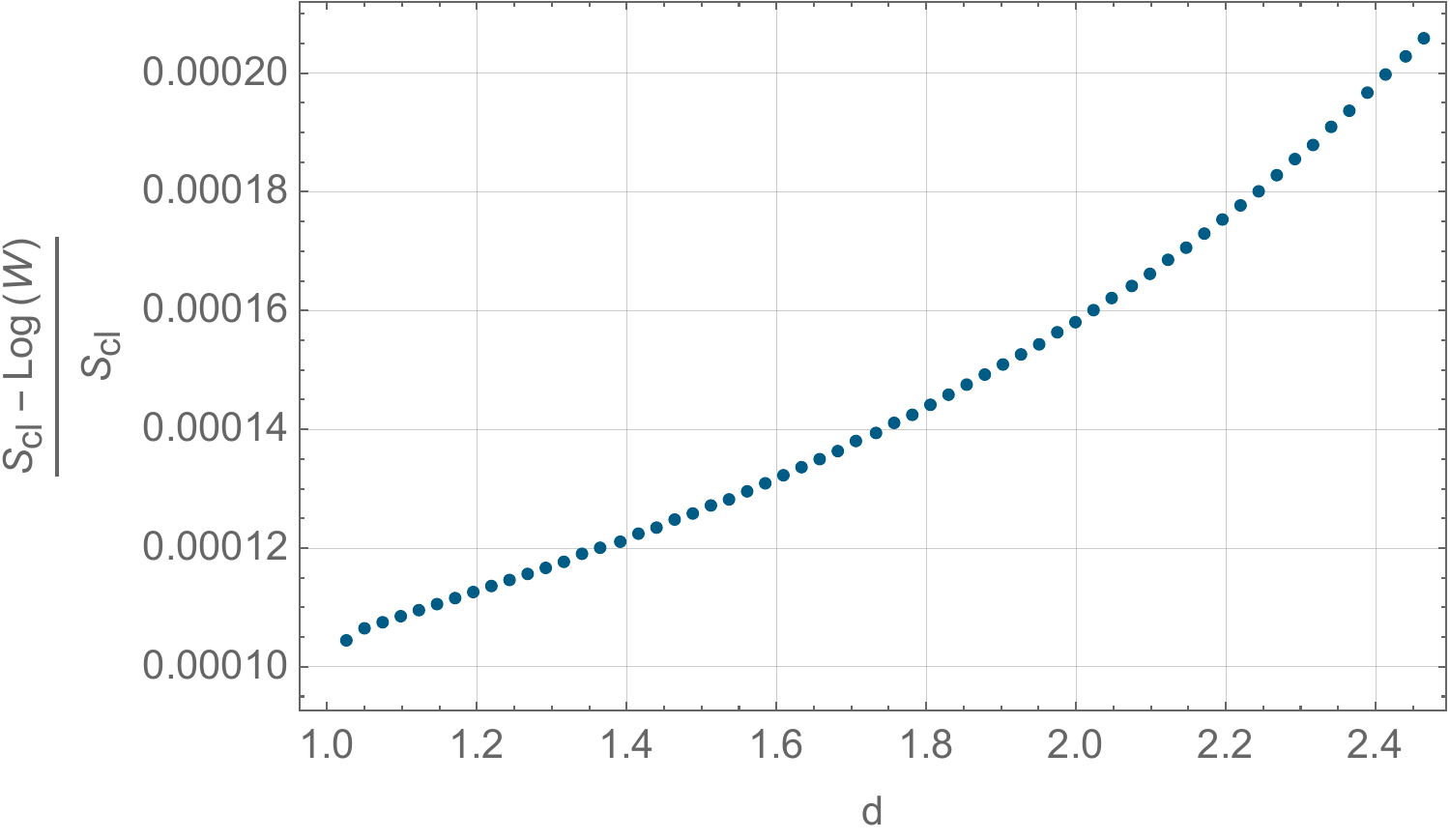}};
	\end{tikzpicture}
	\caption{The relative error for the holographic calculation of the Wilson loop vev plotted as a function of the dimension $d$. We find very good agreement between the analytic expression in \eqref{eq:Wilson-Exact} and the numerical results for $d\lesssim 2.5$. Although the error seems to grow with $d$, this is simply because in producing the data we kept the UV cut-off constant. A closer look at the integrands shows that the fall off of the sub-leading terms grows with $d$ so that the error could be kept constant by gradually increasing the cut-off.}
	\label{fig:WL-fit}
\end{figure}

Another observable of interest to us is the vacuum expectation value of the $\f12$-BPS Wilson loop in the fundamental representation of the gauge group wrapping the equator of the $d$-sphere. This expectation value can be computed holographically by evaluating the regularised on-shell action of a probe string wrapping the equator of the $S^d$ in the spherical brane solutions, i.e. $\log\vev W = -S_{\rm string}^{\rm ren.}$. As shown in \cite{Bobev:2019bvq}, this classical string action for all spherical brane solutions can be expressed as,
\begin{equation}
	S_{\rm string} = \f{1}{\ell_s^2} \int \f{\dd X}{X^\prime(\rho)} \e^{\eta+A}\,. 
\end{equation}
This on-shell action is divergent. In order to obtain a finite answer we need to regularise this expression by adding the counterterm,
\begin{equation}
	S_{\rm counterterm} = \f{1}{g \ell_s^2} \e^{A+\f{3}{7-d}\eta}\Big|_{\rm UV}\,.
\end{equation}
Unfortunately, unlike the supergravity action, the string on-shell action cannot be expressed as a total derivative. Therefore, we need to explicitly perform the numerical integration from the IR to the UV region and cannot simply rely on the IR and UV expansions of the BPS equations. Nevertheless, these numerical integrals can be performed with reasonable accuracy. Importantly, we find that the numerical values for the regularised on-shell action are in excellent agreement with the analytic formula
\begin{equation}\label{eq:Wilson-Exact}
	\log \vev W = 2\pi (4\pi)^{\f{d+1}{2(d-6)}}\left[ 32 \lambda\, \Gamma\left( \f{8-d}{2} \right) \Gamma\left( \f{6-d}{2} \right) \Gamma\left( \f{d-1}{2} \right) \right]^{\f{1}{6-d}}\,.
\end{equation}
The relative error between this analytic expression and the numerical data is presented in Figure \ref{fig:WL-fit}. It is clear that the numerical accuracy in this calculation is worse than the one for the supergravity on-shell action. We attribute this to the necessity of performing two numerical integrations - one to find the numerical supergravity solution and the other to evaluate the on-shell action of the probe string. Nevertheless, we think that our results provide convincing evidence for the validity of the analytic expression in \eqref{eq:Wilson-Exact} for general $d$. Importantly, this holographic result for the Wilson loop vev is in perfect agreement with the results in Section~\ref{sec:BMN} from the supersymmetric localisation matrix model in the dual QFT. This is yet another confirmation of the validity of our supergravity analysis. Moreover, we emphasise that the calculation leading to the holographic result in \eqref{eq:Wilson-Exact} does not rely on any finite counterterms and is thus on a firm footing.

%%%%%%%%%%%%%%%%%
%%%%%%%%%%%%%%%%%

%%%%%%%%%%%%%%%%%%%%%%%%%%%
\section{Thermal solutions}
\label{sec:thermal}
%%%%%%%%%%%%%%%%%%%%%%%%%%%

To further corroborate the validity of our analytic continuation in dimension we study a different set of observables. As discussed in \cite{Bobev:2018ugk} the gauged supergravity truncation summarised in Appendix~\ref{app:Spherical-p} admits a simple non-supersymmetric solution which uplifts to the non-extremal thermal D$p$-brane solutions of 10d supergravity. Here we summarise these solutions and some of their properties with a particular emphasis on the fact that many of the calculations can be done keeping the parameter $d$ general. Furthermore, we leverage the techniques developed in this work to evaluate the on-shell action and vacuum expectation value of a Polyakov loop in these backgrounds. We emphasise that, unlike in the rest of this work, the worldvolume of the D$p$-branes discussed in this section is not spherical but flat, i.e. $S^1 \times \mathbf{R}^p$.

The thermal solutions of interest have constant scalars $X=1$ and $Y=0$ and the metric and dilaton take the form
\begin{equation}
\dd s_{d+1}^2 = \e^{-\eta}\Big( \cD^{1/2}\cH^{-1}\dd r^2 + \cD^{-1/2}(-\cH\dd t^2+\dd x_a\dd x^a) \Big)\,,\qquad \e^{\f{2(d-1)(8-d)}{(7-d)(4-d)}\eta} = \cD\,,
\end{equation}
where the index $a=1,\cdots,p=d-1$ runs over the flat spatial directions on the worldvolume of the brane and the two functions $\cD$ and $\cH$ are given by
\begin{equation}
\cD = \left(gr\right)^{d-8}\,,\qquad 
\cH = 1-\left(\f{r_h}{r}\right)^{8-d}\,.
\end{equation}
Notice that we have chosen coordinates such that the solution uplifts directly to the well-known non-extremal $p$-brane solution as presented in e.g. \cite{Peet:2000hn}. The supersymmetric, extremal solution is found by setting $r_h=0$ but here we will keep $r_h$ general in order to discuss some of its thermal properties. We can calculate the temperature of the black brane by the standard trick of Wick rotating the metric to Euclidean signature and ensuring that the solution caps of smoothly in the IR region $r\to r_h$. This regularity condition leads to the following periodicity of the thermal cycle
\begin{equation}
\beta = \f{4\pi r_h}{(8-d)(g r_h)^{(8-d)/2}}\,,
\end{equation}
with $\beta$ the inverse temperature $T=1/\beta$.

Having introduced the thermal solutions we proceed with the calculation of thermal observables for these supergravity backgrounds. It should be noted that non-extremal $p$-branes exhibit Gregory-Laflamme instability at sufficiently high temperature and therefore our computation is only valid for temperatures sufficiently close to extremality, see \cite{Reall:2001ag}. We start by computing the Euclidean on-shell action. As discussed in Section~\ref{sec:holofreeenergy}, the on-shell supergravity Lagrangian can be reduced to a total derivative, see \eqref{eq:Stotder}, which means that the on-shell action is easily evaluated and reads
\begin{equation}\label{eq:OSABB}
S = \f{\bU_p}{2\kappa_{d+1}^2}\f{g(gr_h)^{8-d}}{T} \left(-2+(10-d)\left(\f{r}{r_h}\right)^{8-d}\right)\bigg|_{r\to\infty}\,,
\end{equation}
where $\bU_p = \int\dvol_p$ denotes the volume of the $p$-dimensional space-like slices of the black brane. Since the spatial slices are non-compact, we will discuss thermodynamic quantities evaluated per unit-volume obtained by dividing by the volume $\bU_p$. Note that we do not get any contribution from the horizon as $r\to r_h$ although the above expression may suggest it. This is because when we write the on-shell action as a total derivative we have integrated in the Gibbons-Hawking boundary term. Effectively this means that the expression in \eqref{eq:OSABB} includes a spurious Gibbons-Hawking term on the horizon which must be subtracted by hand. It so happens that this spurious GH term is the entire contribution as $r\to r_h$. As usual, the on-shell action suffers from UV divergences that need to be regulated by appropriate counterterms. Since the scalar $Y$ vanishes and the UV boundary is flat we only need the counterterm in \eqref{eq:Ssuperpot} which for these simple black brane solutions reduces to a cosmological-constant-like term on the boundary modified by a power of the dilaton. Evaluating this counterterm for the black brane solution we find
\begin{equation}
S_\text{superpot} = -\f{\bU_p}{2\kappa_{d+1}^2}\f{g(gr)^{8-d}}{T}(10-d) \sqrt{1-\left(\f{r_h}{r}\right)^{8-d}}\,\bigg|_{r\to\infty}\,.
\end{equation}
Combining the divergent on-shell action with the above counterterm we obtain the finite expression
\begin{equation}
S_{\rm ren} = \f{\bU_p}{2\kappa_{d+1}^2}\f{g(gr_h)^{8-d}}{2T} (6-d)\,.
\end{equation}
The thermal free energy is related to the Euclidean on-shell action via $F_d = -TS_{\rm ren}$.\footnote{We have added the minus sign by hand since in our conventions for the Euclidean supergravity action it is not already included.} It is easy to verify that the black brane entropy $\cS$ computed as the area of the black brane horizon at $r=r_h$ precisely agrees with the one computed using the regularised on-shell action and the standard thermodynamic relation $\partial F_d/\partial T +\mathcal{S} = 0$. Note that for $d=3$ and $d=4$ our results agrees with Equations (2.8) and (2.10) in \cite{Mateos:2007vn}, while for $d=2$ we find the same results as the one in \cite{Kanitscheider:2008kd}. This serves as a consistency check on our approach and shows that our results can be trusted for general values of $d$. 

We can now rewrite the holographic thermal free energy in terms of the field theory parameters where the dimensionless 't Hooft coupling is defined using the temperature $T$ as the energy scale. We find the following expression 
\begin{equation}\label{eq:Fthermal}
\f{F_d}{\bU_p} = \f{N^2(d-6)}{\Gamma\left(\tfrac{10-d}{2}\right)}\left( \f{16\pi^2}{(d-8)^2}\Gamma\left(\tfrac{8-d}{2}\right)\right)^{\f{8-d}{6-d}} 64^{\f{1}{d-6}}\pi^{\f{d}{d-6}}T^d\lambda_T^{\f{4-d}{d-6}}\,,
\end{equation}
where we have used \eqref{eq:kappadef} and $\lambda_T$ is defined as 
\begin{equation}
	\lambda_T = g_{\rm YM}^2 N T^{d-4}\,.
\end{equation}
Note that this definition of the 't Hooft coupling uses the temperature as the dimensionful parameter and differs from the 't Hooft coupling defined in \eqref{eq:lambdaQFTdef} which uses the radius of $S^d$ to set the scale in the problem and has some fixed numerical coefficients suggested by the supersymmetric localisation matrix model.

In a similar fashion we can compute the expectation value of the Polyakov loop, i.e. a Wilson loop wrapping the thermal circle in the Euclidean supergravity background. For simplicity we will assume that the fundamental string dual to the Polyakov loop takes the same position on the internal space in the 10d uplift of the supergravity solution as the one for the supersymmetric Wilson loop discussed in Section~\ref{subsec:WLvev}. This implies that we are studying a slightly exotic version of the Polyakov loop that has a specific coupling to the scalars in the vector multiplet of the MSYM theory. With these assumptions in mind, a straightforward computation of the regularised on-shell action of the probe string, similar to the one in Section~\ref{subsec:WLvev}, leads to the following holographic result for the Polyakov loop vev for $d\le6$
\begin{equation}\label{eq:PolL}
\log\langle {\cal W}\rangle = -2 \pi \left[\lambda_T \f{2^{7-d}\pi^{\f{4-d}{2}} }{(8-d)^2}\Gamma\Big(\f{8-d}{2}\Big)\right]^{\f{1}{6-d}}\,.
\end{equation}
Importantly, both the thermal free energy in \eqref{eq:Fthermal} and the Polyakov loop vev in \eqref{eq:PolL} exhibit the scaling with the 't Hooft coupling $\lambda$ and the temperature $T$ expected from the scaling similarity discussed in \cite{Biggs:2023sqw}. In addition, the free energy reproduces the correct scaling for the BMN and BFSS matrix quantum mechanics in the $d \rightarrow 1$ limit. This provides further evidence for the validity of our supergravity and holographic calculations for general values of $d$.  Of course, it will be very interesting to calculate these thermal observables by QFT methods and match also the precise numerical coefficients. Unfortunately such a calculation in a thermal strongly interacting QFT is currently out of reach.
 
Finally, let us briefly discuss the supersymmetric and extremal limit of the black D$p$-branes obtained by setting $r_h=0$. In this limit the metric and the dilaton simplify to
\begin{equation}
\dd s_{d+1}^2 = \e^{-\eta}\Big( (gr)^{\f{d-8}{2}}\dd r^2 + (gr)^{\f{8-d}{2}}(-\dd t^2+\dd x_a\dd x^a) \Big)\,,\qquad \e^{\f{2(d-1)(8-d)}{(7-d)(4-d)}\eta} = (gr)^{d-8}\,,
\end{equation}
in particular for $d \to 1$ we obtain
\begin{equation}
\dd s_{2}^2 = \e^{-\eta}\Big( (gr)^{-\f{7}{2}}\dd r^2 - (gr)^{\f{7}{2}}\dd t^2 \Big)\,,\qquad \e^{-\f{14\varepsilon}{9}\eta} = (gr)^{-7}\,,
\end{equation}
where $\varepsilon=(1-d)/2$. As discussed in Section~\ref{sec:dual}, in order to compare this result to the 2d gauged supergravity consistent truncation we must first change variables to $\rho = \e^{\varepsilon \eta}$ which leads to
\begin{equation}
\rho = (gr)^{9/2}\,.
\end{equation}
Furthermore the metric should also be rescaled by a power of the dilaton
\begin{equation}
\dd s_{2,\text{2D sugra}}^2 = \rho^{7/9}\e^\eta\dd s_2^2 =  \Big( \dd r^2 - (gr)^7\dd t^2 \Big)\,.
\end{equation}
We therefore find that, as expected, the $d \to 1$ limit of the extremal black D$p$-brane solutions agrees with the D0-brane solution presented in \eqref{eq:D0solAppA}.

%%%%%%%%%%%%%%%%%%%%%%%%%%%%%%%%%
\section{Discussion and outlook}         
\label{sec:conclusion}	    
%%%%%%%%%%%%%%%%%%%%%%%%%%%%%%%%%

In this paper we studied the large $N$ limit of the $\SU(N)$ MSYM theory on $S^d$ from the perspective of supersymmetric localisation and holography with many of our calculations valid for general values of $d$. We established a precise agreement between the calculation of the free energy on $S^d$ and the vev of a supersymmetric circular Wilson loop at strong coupling and their holographic counterparts computed using the spherical brane supergravity solutions. The $d\to 1$ limit of this analysis leads to the physics of circular D0-branes and bears a resemblance, as well as some marked differences, to the BMN matrix quantum mechanics which we discussed in some detail. Our work leaves a number of open questions and points to some possible generalisations which we now briefly discuss.

\begin{itemize}

\item The supersymmetric localisation matrix model we used to obtain our QFT results was properly derived by studying the MSYM theory on $S^d$ for integer values of $d$ in the range $3 <d < 7$. It will be most interesting to understand whether one can first fix $d=1$ and then perform supersymmetric localisation to arrive at the results we obtained by analytic continuation. Indeed, supersymmetric localisation for the BMN matrix quantum mechanics was studied in \cite{Asano:2012zt,Asano:2014eca} and it will be most interesting to make a connection between this work and our results. By the same token, it will be very interesting to find a relation between the supersymmetric index of the BMN quantum mechanics studied recently in \cite{Chang:2024lkw} and the $S^1$ free energy we computed here.

\item The analysis of supersymmetric vacua for the MSYM theory on $S^d$ in Section~\ref{subsec:vacua} clearly shows that when the gauge group is $\SU(N)$ there are no supersymmetric vacua apart from the trivial one where all scalars vanish. The situation appears to be different if one considers non-compact gauge groups like $\SU(1,N)$ since in this case there are non-trivial embeddings of $\mathfrak{su}(1,1)$ in the gauge group and therefore non-trivial solutions of the supersymmetric vacuum equations in \eqref{eq:Sdsusyvac}. It is important to understand the physics of this further and to study whether such MSYM theories with non-compact gauge groups and the corresponding supersymmetric vacua can be realised by D-branes in string theory.

\item The analysis of \cite{Blau:2000xg}, summarised in Section~\ref{subsec:vacua}, shows that there are MSYM theories on the hyperbolic space $\mathbf{H}_d$. This strongly suggests that there is a realisation of these gauge theories on the worldvolume of D-branes. It will be nice to show this explicitly and to construct supergravity solutions describing the backreaction of these ``hyperbolic branes'' in type II supergravity. Perhaps this can be achieved by following the same approach as the one for the spherical brane solutions in \cite{Bobev:2018ugk,Bobev:2019bvq}. 

\item We used a particular truncation of the 2d $\so(9)$ gauged supergravity presented in \cite{Nicolai:2000zt,Ortiz:2012ib} to construct the circular D0-brane solutions of interest in this work. It is natural to expect that this supergravity theory contain many other supersymmetric solutions of relevance to holography. In particular, it will be most interesting to understand whether there are any supersymmetric black hole solutions with regular horizons in this theory and whether some of the recent lessons learned in the context of JT gravity can be applied to this 2d gravitational model with no AdS$_2$ vacua. 

\item We focused on SYM theories on $S^d$ that preserve the maximal number of 16 supercharges. It is natural to expect that this could be generalised to theories with reduced supersymmetry. One could for instance consider adding mass terms to the MSYM Lagrangian in \cite{Blau:2000xg} that partially breaks supersymmetry and systematically studying the possible supersymmetric Lagrangians. It will be interesting to pursue this further and understand whether, in the spirit of our analysis here, supersymmetric localisation and holographic can be successfully applied to these less supersymmetric setups. 

\item Finally, we note that it will be very interesting to understand further the analytic continuation of our results to $d=0$. This should lead to a deformation of the IKKT matrix model arising on the worldvolume of D$(-1)$-branes which will be very interesting to explore further. Recently, precisely this type of deformations of IKKT that preserve maximal supersymmetry but break the $\so(10)$ R-symmetry to $\su(2)\times \so(7)$ were studied in \cite{Hartnoll:2024csr,Komatsu:2024bop} and it is desirable to understand whether a precise connection with our approach can be established.

\end{itemize}

%%%%%%%%%%%%%%%%%%%%%%%%%%%%%%%%%%%%%
%	ACKNOWLEDGEMENTS
%%%%%%%%%%%%%%%%%%%%%%%%%%%%%%%%%%%%%

\bigskip
\bigskip
\leftline{\bf Acknowledgements}
\smallskip

\noindent 
We acknowledge useful discussions with Ofer Aharony, Shota Komatsu, Ioannis Matthaiakakis, Hynek Paul, Jo\~ao Penedones, Silviu Pufu, Valentina Puletti, and Antoine Vuignier. In particular, we are grateful to Joe Minahan and Anton Nedelin for the collaboration on \cite{Bobev:2019bvq} and especially to Juan Maldacena and Jorge Santos for the inspiring discussions that initiated this project. 
NB is supported in part by FWO projects G003523N, G094523N, and G0E2723N, as well as by the Odysseus grant G0F9516N from the FWO. 
The contributions of PB were made possible through the support of grant No. 494786 from the Simons Foundation (Simons Collaboration on the Non-perturbative Bootstrap) and the ERC Consolidator Grant No. 864828, titled “Algebraic Foundations of Supersymmetric Quantum Field Theory” (SCFTAlg).
FFG is supported by the Icelandic Research Fund under grant 228952-053 and by grants from the University of Iceland Research Fund. 
NB would like to thank the Institute for Advanced Study for the warm hospitality and inspiring atmosphere while part of this project was being completed.

%----------------------------------------------------------------------------------------
%	APPENDICES
%\newpage
\appendix
%----------------------------------------------------------------------------------------

%%%%%%%%%%%%%%%%%%%%%%%%%%%%%%%%%
\section{2d \texorpdfstring{$\so(9)$}{so(9)} gauged supergravity}
\label{app:2dSUGRA}
%%%%%%%%%%%%%%%%%%%%%%%%%%%%%%%%%
	
The three-scalar 2d gravity model studied in the main text can be obtained as a consistent truncation of the maximal $\so(9)$ gauged supergravity presented in \cite{Nicolai:2000zt,Ortiz:2012ib}. This maximal supergravity theory arises as a consistent truncation of type IIA supergravity on $S^8$, see \cite{Bossard:2022wvi} for a recent discussion. The bosonic sector of the maximal theory consists of the metric, a dilaton $\rho$, which is the only $\so(9)$ singlet and couples non-minimally to the metric, 128 additional scalars transforming in the $\bf{84}+\bf{44}$ of $\so(9)$ and $36$ gauge fields transforming in the adjoint of $\so(9)$. 

To capture the physics of the BMN matrix quantum mechanics using 2d supergravity we can perform a further sub-truncation specialising to the $\su(2)\times\so(6)$ invariant sector of the maximal supergravity theory. Decomposing the $\so(9)$ representations under $\su(2)\times\so(6)$ one finds that the only singlets, apart from the metric and dilaton, are one scalar $x$ from the $\bf{44}$ and one pseudo-scalar $\chi$ from the $\bf{84}$.%
\footnote{In the notation of \cite{Ortiz:2012ib}, these scalars are given by
\begin{equation}
	\phi^{789} = \chi \,,\qquad \cV_m{}^a = \diag \left( \e^x,\e^x,\e^x,\e^x,\e^x,\e^x,\e^{-2x},\e^{-2x},\e^{-2x}\right)\,,
\end{equation}
where $\phi^{klm}$ denotes the scalars in the $\bf{84}$ while the scalars in the $\bf{44}$ are collected in the matrix $\cV_{m}{}^a$ and parametrise an $\SL(9)/\SO(9)$ coset.  
} 
Combining the results of \cite{Ortiz:2012ib,Anabalon:2013zka,Ortiz:2014aja} we find the following action for this three-scalar 2d gravity model, 
\begin{equation}
	\label{eq:2daction}
	S = \f{1}{2\kappa_2^2} \int \dvol_{2}\, \rho \left( R-\f92 |\dd x|^2 - 2 \rho^{-2/3}\e^{-6x}|\dd \chi|^2 - V \right)\,,
\end{equation}
where $\dvol_{2}$ is the two-dimensional volume form and the potential $V$ is given by
\begin{equation}
	V= -\f32 \f{g^2}{\rho^{4/9}\e^{2x}}\left[ \left( 8 + 12 \e^{3x} + \e^{6x} \right) +  \rho^{-2/3}\chi^2 \right]\,.
\end{equation}
Here $g$ is the 2d gauge coupling constant which is related to the type IIA string length and the number of D0-branes as $(2\pi \ell_s g)^{-7} = \f{15N}{32\pi^4}$.  It proves useful to write the scalar kinetic terms using the quantity
\begin{equation}
	P_\mu = \partial_\mu x + \f{\ii}{3\rho\e^{3x}}\partial_\mu \chi\,,
\end{equation}
and introduce the superpotential,
\begin{equation}
	W = \f{g}{\rho^{5/9}\e^x}\left( (2+ \e^{3x})\rho^{1/3} + \ii \chi \right)\,,
\end{equation}
in terms of which we can rewrite the potential as
\begin{equation}
	V = \f12 \left( |\partial_x W|^2 + 3 \chi^2 |\partial_\chi W|^2 - 7 |W|^2 \right)\,.
\end{equation}
Note that for $x = \chi = 0$ we obtain a model of 2d dilaton-gravity that is different from the JT gravity model. For instance, there is no $\AdS_2$ vacuum solution. Instead, the supersymmetric $\so(9)$ invariant vacuum solution has a running dilaton and corresponds to the dimensional reduction of the flat D0-brane solution of type IIA supergravity, see \eqref{eq:BPS2dXY0} below.

Instead of the flat, maximally symmetric, Lorentzian, D0-brane solution we are interested in the Euclidean D0-brane solution with a compact $S^1$ worldvolume and the maximal symmetry broken to $\su(1,1)\times\so(6)$. Note that after analytically continuing to Euclidean signature, the R-symmetry is non-compact and the scalar $\chi$ becomes purely imaginary. As argued in the main text, this 2d setup should describe the holographic dual to the Euclidean BMN matrix quantum mechanics on $S^1$. The three supergravity scalars are dual to the three relevant operators in the matrix quantum mechanics. The dilaton $\rho$ is dual to the gauge coupling, while the the scalar $x$ and $\chi$ are dual to the operators in \eqref{eq:dual-ops}.

This type of gauged supergravity consistent truncation is very similar to the supergravity theories used to study spherical brane solutions, see \cite{Bobev:2018ugk}, and in the following we will apply similar methods to analyse the solutions of this theory. In particular, we expect all three scalar to have non-trivial profiles in the bulk.  To construct the solutions we consider the following Euclidean ansatz for the metric 
\begin{equation}
	\ds^2 = \dd r^2 + \cR^2\e^{2A(r)}\dd\tau^2\,,
\end{equation}
where $\tau\sim \tau + 2\pi$ is periodic and the metric function $A$ only depends on the radial coordinate $r$. The constant $\mathcal{R}$ is a placeholder that parametrises the radius of the circle. Assuming that all scalars only depend on the coordinate $r$ and using the supersymmetry variations presented in \cite{Ortiz:2012ib}, we can derive the following system of BPS equations
\begin{equation}\label{BPSeqappendixr}
	\begin{aligned}
		\left(\f{\dd\rho}{\dd r}\right)^2 =&\, \f94 \rho^2 \wti W W\,,\\
		P_r \f{\dd\rho}{\dd r} =&\, -\f12 \rho W\partial_x \wti W\,,\\
		\wti P_r \f{\dd\rho}{\dd r} =&\, -\f12 \rho \wti W\partial_x W\,,\\
		\f{\dd\rho}{\dd r} \f{\dd A}{\dd r} =&\, \f94 \rho^2 W\partial_\rho \wti W + \f94 \rho  \wti W W\,,
	\end{aligned}
\end{equation}
where $\wti W$ is the analytic continuation of $\bar W$ but in Euclidean signature is not necessarily equal to the complex conjugate. It is straightforward to verify that any solution to these equations indeed solves the equations of motion derived from \eqref{eq:2daction}. In the main text it proves useful to redefine the scalar fields as
\begin{equation}\label{eq:xtoX}
%	X = \e^{3x}\,,\quad \text{and}\quad Y = -\ii \rho^{-1/3}\chi\,, 
x= \f13 \log X\,,\quad\text{and}\quad \chi = \ii \rho^{1/3} Y\,,
\end{equation}
and treat $X$ as the radial coordinate. After this rewriting, the BPS equations reduce to the ones in \eqref{eq:BPS0}. These equations can then be used to solve for the metric function $A$ assuming that $Y$ does not vanish. In the remainder of this appendix we explore solutions to the BPS equations where the scalar $Y$ does vanish.
% %
% \begin{equation}
% 	\label{eq:BPS0-appendix}
% 	\begin{aligned}
% 		\f{\dd Y}{\dd X} =&\, -\f{Y}{2X}\f{4+4X+7X^2-Y^2}{4-2X-2X^2-Y^2} \,, \\
% 		\f{\dd \rho}{\dd X} =&\, \f{3\rho}{2X}\f{(2+X)^2-Y^2}{4-2X-2X^2-Y^2}\,,  \\
% 		\f{\dd A}{\dd X} =&\, \f{7}{2X}\f{(2+X)^2-4Y^2}{4-2X-2X^2-Y^2} \,,\\
% 		\f{\dd X}{\dd r} =&\, g \f{(2-2X+Y)}{9X^{4/3}\rho^{2/9}} \,.
% 	\end{aligned}	
% \end{equation}
% %
% When $Y\neq 0$, the metric function $A$ can be found analytically resulting in the following closed form expression,
% %
% \begin{equation}
% 	\e^{2A} = \f{X^{2/3}\rho^{4/9}}{g^2\cR^2Y^2}\left( (2+X)^2 - Y^2 \right)\,,
% \end{equation}
% %
% where the factor of $\cR^2$ in the denominator is due to a convenient choice of an integration constant. After this reparametrization the metric can be written as
% %
% \begin{equation}
% 	\ds^2 = \f{\rho^{4/9}((2+X)^2-Y^2)}{g^2 X^{4/3}}\left( \f{\dd X^2}{(4-2X-2X^2-Y^2)^2} + \f{X^2}{Y^2}\dd\tau^2 \right)\,.
% \end{equation}
% %
% Therefore, the system of BPS equations has been reduced to solving a non-linear differential equation for the function $Y(X)$. Once an expression for $Y(X)$ has been obtained, the dilaton $\rho$ can be found by integrating the second equation in \eqref{eq:BPS0-appendix}. We do not know the general analytic solution for the function $Y(X)$ and in the main text we resort to finding approximate solutions in the UV and IR region as well as numerical integration.

%%%%%%%%%%%
\subsection{Analytic solutions}
%%%%%%%%%%%

The system of 2d supergravity BPS equations admits two simple analytic solutions with $Y=0$. The first is obtained by setting $X=1$ and $Y=0$. In this case the BPS equations simplify to
\begin{equation}\label{eq:BPS2dXY0}
\frac{d\rho}{dr} = \frac{9g}{2} \rho^{7/9}\,,\qquad \frac{dA}{d\rho} = \frac{7}{9\rho}\,.
\end{equation}
The solution to this system of equations is
\begin{equation}\label{eq:D0solAppA}
A = C_1 + \frac{7}{9} \log \rho\,, \qquad \rho = (g r)^{9/2}\,,
\end{equation}
where $C_1$ is an integration constant. This is nothing but the 2d supergravity incarnation of the well-known solution describing the near horizon limit of coincident D0-branes. Indeed, one can check that the 10d uplift of this 2d solution reduces to the background discussed in \cite{Itzhaki:1998dd}. See also Section \ref{sec:thermal} for a discussion of this solution in relation to the thermal black brane backgrounds.

A second class of interesting solutions was constructed in \cite{Ortiz:2012ib} and has two running scalars but a vanishing axion, $\chi$. To obtain this solution in our conventions we set $Y=0$ and study the BPS equations in~\eqref{BPSeqappendixr}. It is also useful to record the BPS equations in terms of the radial variable $r$. They read
\begin{equation}\label{eq:BPSFFGr}
\begin{split}
\frac{d \rho}{d r} &= \frac{3\rho}{2} W\,, \\
\frac{d x}{d r} &= - \frac{1}{3} \partial_{x}W = - \frac{2g}{3} \rho^{-2/9}\e^{-x}(\e^{3x}-1) \,.
\end{split}
\end{equation}
These equations can be readily solved and the solution reads
\begin{equation}\label{eq:solFFG}
\rho = R_0 \frac{X^{3/2}}{(X-1)^{9/4}}\,, \qquad e^{2A} = A_0 \frac{X^{7/3}}{(X-1)^{7/2}}\,,
\end{equation}
where $A_0$ and $R_0$ are integration constants. Note that in writing this solution we have assumed that $X>1$, i.e. $x>0$. One can also choose the other branch, i.e. $x<0$, by appropriately modifying the integration constants $A_0$ and $R_0$. Using the second equation in \eqref{eq:BPSFFGr} and the explicit solution for $\rho(X)$ one finds the 2d metric
\begin{equation}\label{eq:solmetFFG}
\dd s^2 = \frac{9 R_0^{4/9}}{4g^2} \frac{\e^{4x}}{(\e^{3x}-1)^3}\dd x^2 + e^{2A(x)} \dd\tau^2\,.
\end{equation}
After setting $A_0=R_0= g=1$ the results in \eqref{eq:solFFG} and \eqref{eq:solmetFFG} agree precisely with the solution given in (3.7) and (3.8) of \cite{Ortiz:2012ib}. This supergravity background should be interpreted as a continuous distribution of D0-branes which is smeared in an $\su(2) \times \so(6)$ invariant way on the internal $S^8$. The solutions are 2d gauged supergravity analogues of the smeared D3-brane solutions studied in \cite{Freedman:1999gk} and should be the holographic dual of the $\su(2) \times \so(6)$ invariant states on the ``Coulomb branch'' of the BFSS quantum mechanics. This interpretation is indeed consistent with the holographic analysis in \cite{Ortiz:2012ib,Ortiz:2014aja} where it was shown that these solution do not have sources for the operators dual to $\rho$ and $x$.

%%%%%%%%%%%%%%%%%%%%%%%%%%%%%%%%%
\section{Spherical branes}       
\label{app:Spherical-p}	    
%%%%%%%%%%%%%%%%%%%%%%%%%%%%%%%%%	

In the main text we argued that the BPS equations and equations of motion for the two-dimensional dilaton-gravity can be obtained from the $d\to1$ limit of the $d+1$-dimensional gauged supergravity models used in \cite{Bobev:2018ugk,Bobev:2019bvq} to construct supergravity solutions describing spherical branes with $S^d$ worldvolume. Here we briefly summarise the relevant aspects of said spherical brane solutions and refer the reader to \cite{Bobev:2018ugk,Bobev:2019bvq} for more details.

%%%%%%%%%%%
\subsection{Gauged supergravity construction}
%%%%%%%%%%%

The spherical D$(d-1)$-brane solutions can be described uniformly as solutions to a three-scalar truncation of the maximal gauged supergravities in $d+1$ dimensions. These theories can be obtained from the reduction of type II supergravity on $S^{9-d}$, analytically continued to Euclidean signature and truncated to the $\su(1,1)\times\so(7-d)$ invariant sector. The relevant truncation consists of the metric, one real scalar $\eta$ and one complex scalar $\tau$ parametrising an $\SL(2)/\SO(2)$ coset. The bosonic action for the relevant (Euclidean) gauged supergravities is\footnote{To conform with the notation in \cite{Bobev:2018ugk}, in this appendix we use $\tau$ to denote a complex scalar field. This should not be confused with the coordinate $\tau$ used in the main text.}
\begin{equation}
	\label{eq:spherical-action}
	S = \frac{1}{2\kappa_{d+1}^2}\int \dvol_{d+1} \left[ R + \f32\f{d-1}{d-7}|\dd \eta|^2 - 2 \cK_{\tau\tilde{\tau}}|\dd\tau|^2 - V \right]+ S_\text{GH}\,,
\end{equation}
where $\dvol_{d+1}$ is the $d+1$-dimensional volume form, the potential $V$ is given below, the K\"ahler potential on the scalar coset is $\cK = -\log \frac{\tau-\tilde{\tau}}{2}$ and the K\"ahler metric is given by $\cK_{\tau\tilde{\tau}} =\partial_{\tau}\partial_{\tilde{\tau}}\cK$. The relation between the 2d Newton constant $\kappa_{d+1}^2$, the gauged supergravity coupling $g$, the 10d string length $\ell_s$ and string coupling $g_s$ is given by

\begin{equation}
	\kappa_{d+1}^2 = \f{(2\pi\ell_s)^8g_s^2}{8\pi}\f{\Gamma\left( \f{10-d}{2} \right)}{\pi^{\f{10-d}{2}}}g^{9-d}\,.
\end{equation}
For $d>4$ the gauge theories on the worldvolume of the branes are IR free, while for $d<4$ they flow to strongly interacting QFTs in the IR. This difference in IR behaviour is reflected in a dichotomy in the detailed description of the supergravity solutions. Since our interest lies in the limit $d\to 1$ we focus on the $d<4$ case and refer the reader to \cite{Bobev:2018ugk,Bobev:2019bvq} for details on the case $d>4$. For $d<4$, we can define the complex superpotentials
\begin{align}
	\cW = -g \e^{\eta/2}\left( 3\tau + \ii (7-d) \e^{-\frac{d-1}{7-d}\eta} \right)\,,\label{eq:superW}\\
 	\wti{\cW} = -g \e^{\eta/2}\left( 3\wti\tau - \ii (7-d) \e^{-\frac{d-1}{7-d}\eta} \right)\,,\label{eq:superWt}
\end{align}
in terms of which the potential takes the form
\begin{equation}
	V = \f12 \e^\cK \left( \f{7-d}{3(d-1)} \partial_\eta \cW\partial_\eta \wti\cW + \f14 \cK^{\tau\tilde\tau} D_\tau \cW D_{\tilde\tau} \wti\cW - \f{d}{2(d-1)} \cW\wti\cW \right)\,,
\end{equation}
with $D_\tau = \partial_\tau + \partial_\tau \cK$ the K\"ahler covariant derivative.

The spherical brane solutions are domain walls of the Euclidean supergravity with the following metric 
\begin{equation}
	\ds_{d+1}^2 = \dd r^2 + \cR^2\e^{2\cA(r)}\dd\Omega_{d}^2\,,
\end{equation}
where $\dd\Omega_d^2$ is the metric on the round unit radius $d$-sphere with volume $\vol_d = 2\pi^{(d+1)/2}/\Gamma\left((d+1)/2\right)$ and $\cR$ is a bookkeeping constant that indicates the size of the sphere.

A useful redefinition of the scalar fields is provided by
\begin{equation}
	\tau = \ii \e^{-\f{d-1}{7-d}\eta}(X+Y)\,,\qquad \wti\tau = -\ii \e^{-\f{d-1}{7-d}\eta}(X-Y)\,.
\end{equation}
In terms of these fields, and using $X$ as the radial coordinate, we can uniformly write the BPS equations for any $d<4$ as
\begin{equation}
	\begin{aligned}
 		Y'(X) &= -\f{Y}{2X}\f{(d-3)^2+4(2-d)X+7X^2-Y^2}{2(1-X)(3-d+X)-Y^2}\,,\\
 		\eta'(X) &= -\f{7-d}{2(d-1) X}\f{(3-d+X)^2-Y^2}{2(1-X)(3-d+X)-Y^2}\,,
	\end{aligned}
\end{equation}
while the BPS equation for the warp factor, $\cA$, reduces to the following algebraic expression
\begin{equation}
	\cR^2\e^{2\cA} = \e^{\f{2(d-4)}{d-7}\eta}X\f{ (3-d+X)^2-Y^2}{g^2 Y^2}\,.
\end{equation}
Finally, using the coordinate $X$ we can write the metric as
\begin{equation}
	\ds_{d+1}^2 = \f{((3-d+X)^2-Y^2)}{X g^2 }\e^{\f{-2(4-d)}{7-d}\eta}\left( \f{\dd X^2}{(2(1-X)(3-d+X)-Y^2)^2} + \f{X^2}{Y^2}\dd\Omega_{d}^2\right)\,.
\end{equation}
For completeness, we note that the derivative of $X$ with respect to $r$ can be written as, 
\begin{equation}
	X^\prime (r) = g\,\e^{\f{d-4}{d-7}\eta} \f{\sqrt{X}(2(d-3+(2-d)X+X^2)+Y^2)}{\sqrt{(3-d+X)^2-Y^2}}\,. 
\end{equation}
%

%%%%%%%%%%%
\subsection{Uplift to ten-dimensional type II supergravity}
%%%%%%%%%%%

Any solution to the above BPS equations can be uplifted to type IIA or IIB supergravity depending on the value of $d$. The type II string frame metric for these backgrounds is given by
\begin{equation}
	\label{eq:spherical-metric}
	\ds^2 = \f{\e^\eta}{\sqrt{Q}}\left[ \ds^2_{d+1} + \f{\e^{\f{2(d-4)}{7-d}\eta}}{g^2}\left( \dd\theta^2+ P\cos^2\theta \dd\wti\Omega_2^2 + Q\sin^2\theta \dd\Omega_{(6-d)}^2 \right)\right]\,,
\end{equation}
where $\dd\wti\Omega_2^2$ is the metric on unit radius $\dS_2$ and $\dd\Omega_n^2$ is the metric on the unit radius round $S^n$ with volume $\vol_n$. The gauged supergravity coupling $g$ can be expressed in terms of string theory quantities as
\begin{equation}
	(2\pi\ell_s g)^{d-8} = \f{g_s N}{2\pi \vol_{7-d}}\,.
\end{equation}
The internal metric is topologically a $\dS_{9-d}$, with the squashing functions $P$ and $Q$ given by
\begin{equation}
	P = \f{X}{X\sin^2\theta + (X^2-Y^2)\cos^2\theta}\,,\qquad\quad Q = \f{X}{\sin^2\theta+ X\cos^2\theta}\,.
\end{equation}
The ten-dimensional dilaton has the following form,
\begin{equation}
	\label{eq:spherical-dilaton}
	\e^{2\Phi} = g_s^2 \e^{\f{(d-1)(8-d)}{7-d}\eta} P Q^{\f{2-d}{2}}\,.
\end{equation}
Finally, the non-trivial form fields in the background are given by
\begin{equation}
	\label{eq:spherical-forms}
	\begin{aligned}
		B_2 =& \e^{\f{d-1}{7-d}\eta}\f{YP}{g^2 X}\cos^3\theta\,\wti\dvol_2\,,\\
		C_{6-d} =& \ii\e^{-\f{d-1}{7-d}\eta}\f{YQ}{g_s g^{6-d} X}\sin^{5-d}\theta\,\dvol_{6-d}\,,\\
		C_{8-d} =& \f{\ii}{g_s g^{8-d}}\left( \omega(\theta) + P \cos\theta\sin^{7-d}\theta \right)\,\wti\dvol_2\wedge\dvol_{6-d}\,,\\
	\end{aligned}
\end{equation}
where $\wti\dvol_2$ and $\dvol_n$ are respectively the volume forms on unit radius $\dS_2$ and $S^n$. The function $\omega(\theta)$ is defined such that in the UV, i.e. when $(X,Y)\to (1,0)$, the exterior derivative of $C_{8-d}$ reduces to a multiple of the volume form of $\dS_{9-d}$,
\begin{equation}
	\f{\dd}{\dd\theta}\left( \omega(\theta) + \cos\theta \sin^{7-d}\theta \right) = (8-d)\cos^2\theta \sin^{6-d}\theta\,.
\end{equation}
The UV region of the spherical brane solutions is located at $(X,Y)=(1,0)$, where the solutions reduce to the flat D$p$-brane solutions of type II supergravity \cite{Itzhaki:1998dd}. The IR region is located at
\begin{equation}
	\label{eq:IR-d}
	X_{\rm IR} = \f{d-1}3\,,\qquad\quad Y_{\rm IR} = \pm \f{2(d-4)}{3}\,,
\end{equation}
where the scalar $\eta$ approaches a constant value $\eta_{IR}$ and the metric smoothly caps off.

In order to compare these solutions with their holographic duals, i.e. the MSYM theory on $S^d$, it will be useful to explicitly provide the holographic dictionary between the various parameters on the gravity and gauge theory side. In our conventions, the D-brane tension and Yang-Mills coupling constants are given in terms of string theory quantities as
\begin{equation}
	\mu_{d-1} = \f{2\pi}{(2\pi\ell_s)^d}\,,\qquad\quad g_{\rm{YM}}^2 = \f{(2\pi)^2g_s}{(2\pi\ell_s)^4\mu_{d-1}} = \f{2\pi g_s}{(2\pi\ell_s)^{4-d}}\,. 
\end{equation}
Finally, following \cite{Bobev:2019bvq}, we define the dimensionless holographic 't~Hooft constant as
\begin{equation}
	\lambda = \f{2\pi g_s N}{(2\pi\ell_s)^{4-d}}\cR^{4-d}\e^{(4-d)\cA}\e^{\f{10-d}{7-d}\eta}\bigg|_{\rm UV}\,.
\end{equation}
%

%----------------------------------------------------------------------------------------
%	REFERENCES
%---------------------------------------------------------------------------------------- 
\newpage
\bibliography{SphericalD0}
\bibliographystyle{JHEP}
	
\end{document}